\documentclass[twocolumn,american,aps,prb]{revtex4}
\usepackage{mathptmx}

\usepackage[T1]{fontenc}
\usepackage[latin9]{inputenc}
\usepackage{float}
\usepackage{amsmath}
\usepackage{amssymb}
\usepackage{graphicx}

\makeatletter
\@ifundefined{textcolor}{}
{%
 \definecolor{BLACK}{gray}{0}
 \definecolor{WHITE}{gray}{1}
 \definecolor{RED}{rgb}{1,0,0}
 \definecolor{GREEN}{rgb}{0,1,0}
 \definecolor{BLUE}{rgb}{0,0,1}
 \definecolor{CYAN}{cmyk}{1,0,0,0}
 \definecolor{MAGENTA}{cmyk}{0,1,0,0}
 \definecolor{YELLOW}{cmyk}{0,0,1,0}
}

\usepackage{float}

\usepackage{textcomp}

\usepackage{esint}



\makeatother

\usepackage{babel}
\begin{document}

\title{Charged-particle oscillation in DC voltage biased plane-parallel
conductors }

\author{Sung Nae Cho }

\email{sungnae.cho@samsung.com}

\selectlanguage{american}%

\affiliation{Micro Devices Group, Micro Systems Laboratory, Samsung Advanced Institute
of Technology, Samsung Electronics Co., Ltd, Mt. 14-1 Nongseo-dong,
Giheung-gu, Yongin-si, Gyeonggi-do 446-712, Republic of Korea. }

\date{16 March 2012 }
\begin{abstract}
\noindent The phenomenon of charged-particle oscillation  in DC voltage
biased plane-parallel conductors is discussed. Traditionally accepted
mechanism for explaining the oscillatory behavior of charged particles
in such system attributes the phenomenon to  a process of charge exchange,
which takes place when charged-particle is in close proximity to one
of the electrodes. A novel finding here reveals that for microscopic
or smaller particles under special circumstances, charged-particle
oscillation does not involve charge exchanges. Such system radiates
and the frequency of emitted radiation is controlled by a DC voltage
biased across the two plane-parallel electrodes.
\end{abstract}
\maketitle
\noindent \textbf{This work has been published in Physics of Plasmas
with open access (author select) option. The reference to the article
is Phys. Plasmas 19, 033506 (2012); http://dx.doi.org/10.1063/1.3690104. }

\section{Introduction }

The phenomenon of charged-particle oscillation in DC voltage biased
plane-parallel conductors (or electrodes) is well known.\cite{Tobazeon-1,Musinski-2,Szirmai-3}
Such phenomenon has been extensively studied over years, both theoretically
and experimentally, due to its usefulness in variety of applications
such as electrostatic thruster and nanoprinting, for instance, which
require highly energetic charged-nanoparticles with very high speed.\cite{Th-Throttenberg-4,Musinski-5,Musinski-6,Musinski-7,Musinski-8} 

\begin{figure}[H]
\begin{centering}
\includegraphics[width=1\columnwidth]{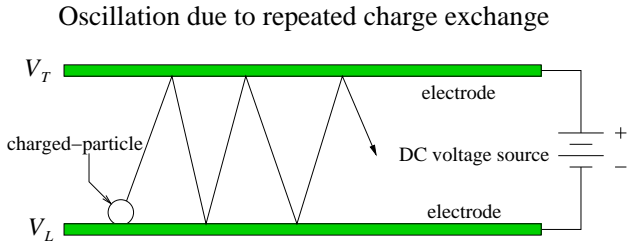}
\par\end{centering}

\caption{(Color online) Schematic of charged-particle oscillation due to repeated
charge exchange. \label{fig:oscillation_due_to_charge_exchange}}
\end{figure}

Traditionally, the phenomenon of charged-particle oscillation subjected
to a constant electric field has been attributed to a process of charge
exchange, which takes place when charged-particle is in close proximity
to one of the electrodes.\cite{Tulagin} When charged-particle is
placed between a DC voltage biased plane-parallel electrodes, it migrates
to the electrode of opposite polarity. For electrically conducting
particles, charge exchange takes place near the point of contact with
the electrode. This reverses the polarity of the charged-particle
and the particle gets repulsed towards the opposite electrode. There,
again, the charge exchange occurs and this process gets repeated,
resulting in charged-particle oscillation between the electrodes.
Such process is schematically illustrated in Fig. \ref{fig:oscillation_due_to_charge_exchange}. 

\begin{figure}[h]
\begin{centering}
\includegraphics[width=1\columnwidth]{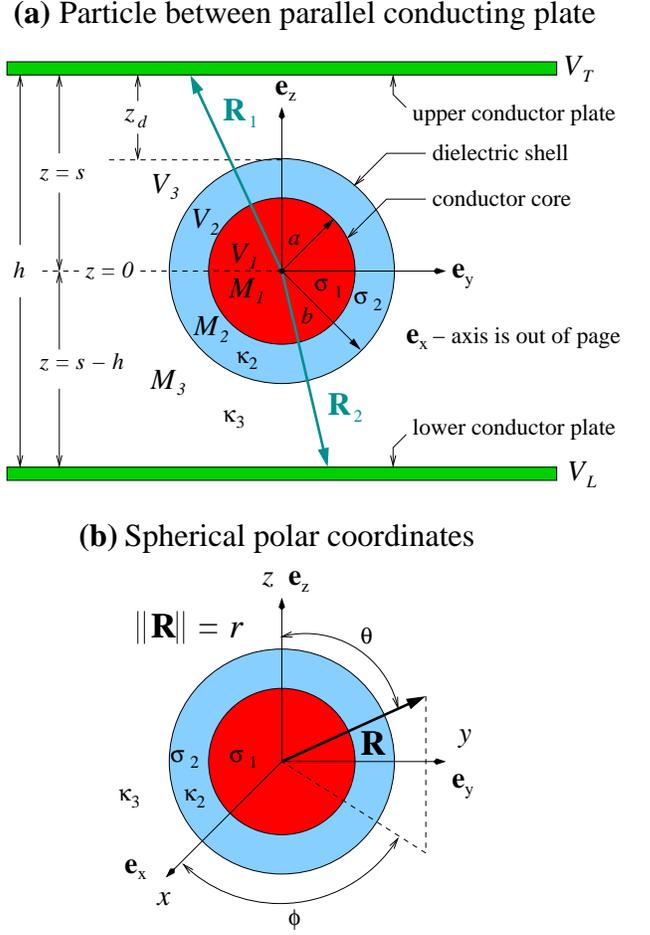}
\par\end{centering}

\caption{(Color online) (a) Cross-sectional view of an ionized core-shell structured
particle confined by the DC voltage biased plane-parallel conductors.
(b) Spherical polar coordinate system showing spherical polar triplet
$\left(r,\theta,\phi\right)$ of a vector $\mathbf{R}$ in Euclidean
three dimensional space, $\mathbb{R}^{3}.$ \label{fig:particle-in-capacitor} }
\end{figure}

A novel finding in this work reveals that under special circumstances,
core-shell structured charged-particle oscillates under constant electric
field. Such oscillatory behavior cannot be explained by the aforementioned
traditional picture because it does not involve any charge exchanges.
The criterion for such oscillatory modes is given by 
\begin{equation}
Q_{T}>\frac{2\pi\epsilon_{0}\kappa_{3}}{\xi}\left(\left|\psi\right|E_{p}+\sqrt{\psi^{2}E_{p}^{2}+\frac{16mg\xi}{\pi\epsilon_{0}\kappa_{3}}}\right),\label{eq:OSC-CRITERION}
\end{equation}
 where 
\[
\xi=\frac{1}{z_{d,m}^{2}}-\frac{1}{\left(h-z_{d,m}\right)^{2}}>0,
\]
 
\[
\psi=\frac{\gamma\left(b^{3}-a^{3}\right)-b^{3}}{z_{d,m}^{3}}+\frac{\gamma\left(b^{3}-a^{3}\right)-b^{3}}{\left(h-z_{d,m}\right)^{3}}-8<0,
\]
 and 
\[
\gamma=\frac{3\kappa_{3}b^{3}}{\left(\kappa_{2}+2\kappa_{3}\right)b^{3}+2\left(\kappa_{2}-\kappa_{3}\right)a^{3}}<1.
\]
 Here, $\epsilon_{0}$ is the permittivity of free space, $g=9.8\,\textnormal{m}\cdot\textnormal{s}^{-2}$
is the gravitational constant, $h$ is the gap between the two plane-parallel
electrodes, $Q_{T}$ is the positive effective charge carried by the
core-shell structured particle, $m$ is the mass of the particle,
and $\kappa_{3}$ is the dielectric constant for the space between
the two electrodes. The core-shell structured particle, which is illustrated
in Fig. \ref{fig:particle-in-capacitor}, has a outer radius of $b,$
the inner conductor core radius of $a,$ and the insulator shell layer
has a dielectric constant of $\kappa_{2}.$ The term $z_{d,m}$ is
the maximum value assumed by the parameter $z_{d}$ in Fig. \ref{fig:particle-in-capacitor}
as the particle executes an oscillatory motion. The particle oscillates
back and forth between $z_{d}=0$ and $z_{d}=z_{d,m};$ hence, physically,
$z_{d,m}$ represents the turning point where the core-shell structured
charged-particle begins to move back towards the conductor plate located
at $z_{d}=0.$ 

For a negatively charged core-shell structured particle, the oscillatory
criterion is given by 
\begin{equation}
\left|Q_{T}\right|>\frac{2\pi\epsilon_{0}\kappa_{3}}{\eta}\left(\left|\psi\right|E_{p}-\sqrt{\psi^{2}E_{p}^{2}-\frac{16mg\eta}{\pi\epsilon_{0}\kappa_{3}}}\right),\label{eq:OSC-CRITERION-nC}
\end{equation}
 where 
\[
\eta=\frac{1}{\left(h-z_{d,m}\right)^{2}}-\frac{1}{z_{d,m}^{2}}>0.
\]
The two criteria, Eqs. (\ref{eq:OSC-CRITERION}) and (\ref{eq:OSC-CRITERION-nC}),
are not form wise identical due to the fact that negatively charged
particle oscillates near the lower conductor plate whereas the positively
charged particle oscillates near the upper conductor plate. Here,
the upper conductor plate, which is located at $z_{d}=0,$ is assumed
to be held at a higher voltage than the lower conductor plate, which
is located at $z_{d}=h.$ The finding in this work reveals that for
the configuration illustrated in Fig. \ref{fig:particle-in-capacitor},
the positively charged core-shell structured particle only oscillates
in the region where $0<z_{d}<h/2;$ and, the negatively charged core-shell
structured particle oscillates only in the region where $h/2<z_{d}<h.$ 

This remarkable result, which I find it quite obvious but not trivial,
is a direct consequence of solving electrostatic boundary value problem
involving core-shell structured charged-particle subjected to DC voltage
biased plane-parallel electrodes depicted in Fig. \ref{fig:particle-in-capacitor}.
The problem is analyzed by first solving the electric potential $V_{3}$
in the region $M_{3}.$ The solution to $V_{3}$ is obtained by solving
Laplace equation with appropriate electrostatic boundary conditions.
Thereafter, induced charges on the surface of each plane-parallel
electrodes are computed. Similarly, the effective charge carried by
the core-shell structured particle is also computed. The dynamics
of the charged-particle is formulated by considering Coulomb forces
between the charged-particle and the induced charges on the surface
of each plane-parallel electrodes. To generalize the problem to include
all particle speed ranges, the dynamics is formulated in relativistic
formalism. 

This article is organized as follow: (I) introduction, (II) outline
of results, (III) theory, (IV) concluding remarks, and (V) acknowledgments.
Normally, the section of theory immediately follows the introduction.
However, as this work involves lengthy derivations, albeit straightforward,
the reader is prone to lose the essence of what this paper tries to
portray. For this reason, the section titled ``outline of results''
is placed immediately after the introduction. In the outline of results,
no details of derivations are provided. Instead, the essence of this
article is briefly summarized there using only the results, which
are rigorously derived in the theory section.

\section{Outline of results}

The essence of this article is to investigate the dynamics of charged-particle
illustrated in Fig. \ref{fig:particle-in-capacitor}(a), where a conducting
sphere coated with a shell of dielectric is placed in an otherwise
constant electric field. The upper conductor plate electrode is held
at a DC voltage of $V_{T}$ and the lower conductor plate electrode
is held at DC voltage of $V_{L},$ where $V_{L}<V_{T}.$ The two plane-parallel
conductor plate electrodes are separated by a gap of $h.$ The conducting
sphere coated with a shell of dielectric has a free charge densities
of $\sigma_{1}$ and $\sigma_{2},$ where $\sigma_{1}$ is the free
charge density on the surface of the conducting sphere and $\sigma_{2}$
is the free charge density on the outermost surface of the dielectric
shell. The free charge on the surface of dielectric shell has been
introduced purely for generalization of the problem. Without loss
of generality, $\sigma_{2}$ can be set to zero in the final form
of solution.

That said, adopting the particle coordinate system illustrated in
Fig. \ref{fig:particle-in-capacitor}(b), the electric potential in
region $M_{3}$ of Fig. \ref{fig:particle-in-capacitor}(a) is given
by 
\begin{align*}
 & V_{3}\left(r,\theta\right)=V_{L}+E_{p}\left(h-s+r\cos\theta\right)+\frac{\nu}{r}\\
 & \quad+\frac{\left[\gamma\left(b^{3}-a^{3}\right)-b^{3}\right]E_{p}\cos\theta}{r^{2}}+C,\quad r>b,
\end{align*}
 where $C$ is a constant, $E_{p}$ is the magnitude of DC electric
field in the gap between two parallel plates in the absence of core-shell
structured charged-particle, 
\begin{align*}
E_{p} & =\frac{1}{h}\left(\left|V_{T}-V_{L}\right|\right);
\end{align*}
 and, $\gamma$ and $\nu$ are defined as 
\begin{equation}
\gamma=\frac{3\kappa_{3}b^{3}}{\left(\kappa_{2}+2\kappa_{3}\right)b^{3}+2\left(\kappa_{2}-\kappa_{3}\right)a^{3}},\label{eq:gama-OR}
\end{equation}
 
\begin{align}
\nu & =\frac{2a\left(b-a\right)\sigma_{1}}{\epsilon_{0}\kappa_{2}}+\frac{a^{2}\sigma_{1}+b^{2}\sigma_{2}}{\epsilon_{0}\kappa_{3}}.\label{eq:nu-OR}
\end{align}
 Here, $\epsilon_{0}$ is the electric permittivity of free space
and $\kappa_{2}$ and $\kappa_{3}$ are dielectric constants respectively
for regions $M_{2}$ and $M_{3}$ illustrated in Fig. \ref{fig:particle-in-capacitor}.
With $V_{3}\left(r,\theta\right)$ and application of appropriate
electric boundary conditions to each conductor plates yields 
\begin{align*}
\sigma_{iup} & =-\epsilon_{0}\kappa_{3}\left\{ \frac{3\left[\gamma\left(b^{3}-a^{3}\right)-b^{3}\right]E_{p}s^{2}}{\left(\rho^{2}+s^{2}\right)^{5/2}}\right.\\
 & \left.+\frac{\nu s-\left[\gamma\left(b^{3}-a^{3}\right)-b^{3}\right]E_{p}}{\left(\rho^{2}+s^{2}\right)^{3/2}}-E_{p}\right\} 
\end{align*}
 and 
\begin{align*}
\sigma_{ilp} & =\epsilon_{0}\kappa_{3}\left\{ \frac{3\left[\gamma\left(b^{3}-a^{3}\right)-b^{3}\right]E_{p}\left(h-s\right)^{2}}{\left[\rho^{2}+\left(h-s\right)^{2}\right]^{5/2}}\right.\\
 & \left.-\frac{\nu\left(h-s\right)+\left[\gamma\left(b^{3}-a^{3}\right)-b^{3}\right]E_{p}}{\left[\rho^{2}+\left(h-s\right)^{2}\right]^{3/2}}-E_{p}\right\} ,
\end{align*}
 where $\sigma_{iup}\equiv\sigma_{iup}\left(\rho,s\right)$ is the
induced surface charge density on the surface of upper conductor plate,
$\sigma_{ilp}\equiv\sigma_{ilp}\left(\rho,s\right)$ is the induced
surface charge density on the surface of lower conductor plate, and
$\rho\equiv\sqrt{x^{2}+y^{2}}.$ 

The net force exerted on the core-shell structured charged-particle
by induced charges on each surfaces of the conductor plates is given
by $\mathbf{F}=\mathbf{F}_{1}+\mathbf{F}_{2},$ 
\begin{align}
\mathbf{F}_{i} & =-\frac{Q_{T}}{8\pi\epsilon_{3}}\int_{\phi_{i}=0}^{2\pi}\int_{\rho_{i}=0}^{\rho}\frac{\varsigma_{i}\mathbf{R}_{i}\rho_{i}d\rho_{i}d\phi_{i}}{\left(\mathbf{R}_{i}\cdot\mathbf{R}_{i}\right)^{3/2}},\label{eq:Fi-OR}
\end{align}
 where $i=\left(1,2\right),$ $\varsigma_{1}\equiv\sigma_{iup},$
$\varsigma_{2}\equiv\sigma_{ilp},$ $\epsilon_{3}$ is the electric
permittivity of the region $M_{3},$ and $\mathbf{R}_{i}$ is given
by 
\begin{align*}
\mathbf{R}_{1} & =\mathbf{e}_{x}\rho_{1}\cos\phi_{1}+\mathbf{e}_{y}\rho_{1}\sin\phi_{1}+\mathbf{e}_{z}s,\\
\mathbf{R}_{2} & =\mathbf{e}_{x}\rho_{2}\cos\phi_{2}+\mathbf{e}_{y}\rho_{2}\sin\phi_{2}+\mathbf{e}_{z}\left(s-h\right).
\end{align*}
 In the limit the charged-particle becomes very small compared to
the dimensions of parallel plates, which is the case for micro- or
nano-sized charged-particle confined in a microscopically large, but
macroscopically small parallel plates, the $\mathbf{F}_{i}$ of Eq.
(\ref{eq:Fi-OR}) for $i=\left(1,2\right)$ can be shown to become
\begin{align}
\mathbf{F}_{1} & =\mathbf{e}_{z}\frac{Q_{T}}{4}\left\{ \frac{\nu}{4s^{2}}+\frac{\left[\gamma\left(b^{3}-a^{3}\right)-b^{3}\right]E_{p}}{4s^{3}}-E_{p}\right\} \label{eq:F1-with-QT}
\end{align}
 and 
\begin{align}
\mathbf{F}_{2} & =\mathbf{e}_{z}\frac{Q_{T}}{4}\left\{ \frac{\left[\gamma\left(b^{3}-a^{3}\right)-b^{3}\right]E_{p}}{4\left(h-s\right)^{3}}-\frac{\nu}{4\left(h-s\right)^{2}}-E_{p}\right\} ,\label{eq:F2-with-QT}
\end{align}
 where 
\begin{align*}
Q_{T} & =8\pi a\left(b-a\right)\sigma_{1}\frac{\kappa_{3}}{\kappa_{2}}+4\pi\left(a^{2}\sigma_{1}+b^{2}\sigma_{2}\right).
\end{align*}
 When the gravitational effect is included, the force experienced
by the core-shell structured charged-particle is 
\begin{align*}
\mathbf{F}_{T} & =\mathbf{F}_{1}+\mathbf{F}_{2}-\mathbf{e}_{z}mg
\end{align*}
 or 
\begin{align}
\mathbf{F}_{T} & =\mathbf{e}_{z}\frac{Q_{T}}{16}\left\{ \frac{\nu}{s^{2}}-\frac{\nu}{\left(h-s\right)^{2}}+\frac{\left[\gamma\left(b^{3}-a^{3}\right)-b^{3}\right]E_{p}}{s^{3}}\right.\nonumber \\
 & \left.+\frac{\left[\gamma\left(b^{3}-a^{3}\right)-b^{3}\right]E_{p}}{\left(h-s\right)^{3}}-8E_{p}\right\} -\mathbf{e}_{z}mg,\label{eq:FT-OR-with-QT}
\end{align}
 where $m$ is the mass of the particle, $g=9.8\,\textnormal{m}\cdot\textnormal{s}^{-2}$
is the gravity constant, and the gravitational force has been assumed
to be in the $-\mathbf{e}_{z}$ direction. It can be shown that $Q_{T}$
is related to $\nu$ by 
\[
Q_{T}=4\pi\epsilon_{0}\kappa_{3}\nu;
\]
 and the force $\mathbf{F}_{T}$ may be re-expressed, for convenience,
as 
\begin{align}
\mathbf{F}_{T} & =\mathbf{e}_{z}\frac{\pi\epsilon_{0}\kappa_{3}\nu}{4}\left\{ \frac{\nu}{s^{2}}-\frac{\nu}{\left(h-s\right)^{2}}+\frac{\left[\gamma\left(b^{3}-a^{3}\right)-b^{3}\right]E_{p}}{s^{3}}\right.\nonumber \\
 & \left.+\frac{\left[\gamma\left(b^{3}-a^{3}\right)-b^{3}\right]E_{p}}{\left(h-s\right)^{3}}-8E_{p}\right\} -\mathbf{e}_{z}mg.\label{eq:FT-OR}
\end{align}
 It is noticed that $\mathbf{F}_{T},$ which is the net force exerted
on the core-shell structured charged-particle illustrated in Fig.
\ref{fig:particle-in-capacitor}(a), is a one dimensional force that
only depends on the relative length, $s,$ measured between the particle
and the surface of the upper parallel plate electrode. 

The dynamics of oscillating charged-particle is given by 

\[
\mathbf{e}_{z}\frac{d}{dt}\left(\frac{mv}{\sqrt{1-\frac{v^{2}}{c^{2}}}}\right)=\mathbf{F}_{T},
\]
 where $c=3\times10^{8}\,\textnormal{m}\cdot\textnormal{s}^{-1}$
is the speed of light in vacuum. Using the explicit expression for
$\mathbf{F}_{T},$ Eq. (\ref{eq:FT-OR}), it can be shown that 
\begin{align*}
\ddot{s} & =\left(1-\frac{\dot{s}^{2}}{c^{2}}\right)^{3/2}\left(\frac{\pi\epsilon_{0}\kappa_{3}\nu}{4m}\left\{ \frac{\nu}{s^{2}}-\frac{\nu}{\left(h-s\right)^{2}}\right.\right.\\
 & +\frac{\left[\gamma\left(b^{3}-a^{3}\right)-b^{3}\right]E_{p}}{s^{3}}\\
 & \left.\left.+\frac{\left[\gamma\left(b^{3}-a^{3}\right)-b^{3}\right]E_{p}}{\left(h-s\right)^{3}}-8E_{p}\right\} -g\right),
\end{align*}
 where $\mathbf{e}_{z}$ has been dropped for convenience and the
notations $\dot{s}$ and $\ddot{s}$ respectively denote the first
and second time derivatives, i.e., $\dot{s}\equiv ds/dt$ and $\ddot{s}\equiv d^{2}s/dt^{2}.$ 

In terms of the $z_{d}$ parameter illustrated in Fig. \ref{fig:particle-in-capacitor}(a),
\[
s=z_{d}+b,\quad\dot{s}=\dot{z}_{d},\quad\ddot{s}=\ddot{z}_{d},
\]
which is the separation length between the upper electrode plate and
the uppermost surface of the core-shell structured charged-particle,
the previous nonlinear ordinary differential equation becomes 
\begin{align}
\ddot{z}_{d} & =\left(1-\frac{\dot{z}_{d}^{2}}{c^{2}}\right)^{3/2}\left(\frac{\pi\epsilon_{0}\kappa_{3}\nu}{4m}\left\{ \frac{\nu}{\left(z_{d}+b\right)^{2}}\right.\right.\nonumber \\
 & -\frac{\nu}{\left(h-z_{d}-b\right)^{2}}+\frac{\left[\gamma\left(b^{3}-a^{3}\right)-b^{3}\right]E_{p}}{\left(z_{d}+b\right)^{3}}\nonumber \\
 & \left.\left.+\frac{\left[\gamma\left(b^{3}-a^{3}\right)-b^{3}\right]E_{p}}{\left(h-z_{d}-b\right)^{3}}-8E_{p}\right\} -g\right).\label{eq:ODE-rela-OR}
\end{align}

\begin{figure}[h]
\begin{centering}
\includegraphics[width=1\columnwidth]{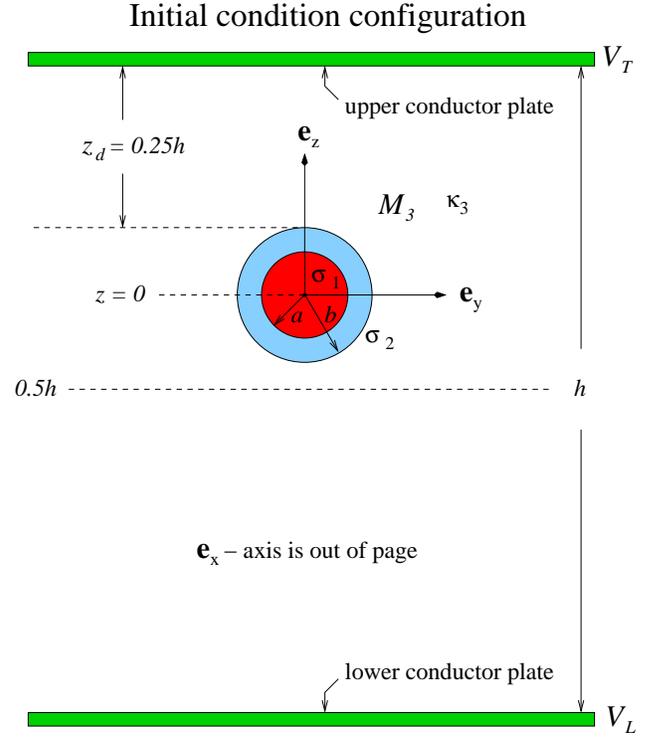}
\par\end{centering}

\caption{(Color online) Initial position of the particle. There are no free
surface charges on the insulating shell; and, hence, $\sigma_{2}=0\,\textnormal{C}\cdot\textnormal{m}^{-2}.$
\label{fig:initial_configuration}}
\end{figure}

To solve and plot Eq. (\ref{eq:ODE-rela-OR}), the core-shell structured
particle in Fig. \ref{fig:particle-in-capacitor} has been chosen
to be the aluminum nanoparticle, where the core is aluminum and the
shell is aluminum oxide. The following parameter values have been
assigned: 
\begin{equation}
\left\{ \begin{array}{c}
\kappa_{2}=6,\quad\kappa_{3}=1,\\
a=1.5\,\mu\textnormal{m},\quad h=1\,\textnormal{mm},\\
b-a=4\,\textnormal{nm},\\
V_{T}=8\,\textnormal{kV},\quad V_{L}=0\,\textnormal{V},\\
\sigma_{1}=0.014\,\textnormal{C}\cdot\textnormal{m}^{-2},\\
\sigma_{2}=0\,\textnormal{C}\cdot\textnormal{m}^{-2},\\
\rho_{m,1}=2700\,\textnormal{kg}\cdot\textnormal{m}^{-3},\\
\rho_{m,2}=3800\,\textnormal{kg}\cdot\textnormal{m}^{-3},
\end{array}\right.\label{eq:parameter_VALUE}
\end{equation}
 where $\rho_{m,1}$ and $\rho_{m,2}$ are mass densities of the aluminum
core and the aluminum oxide, respectively. The thickness of aluminum
oxide layer has been set at $4\,\textnormal{nm},$ which is typical
of aluminum nanoparticles.\cite{Japan-al2o3} Because aluminum oxide
is an high-k dielectric material, i.e., $k_{2}\sim6,$ the $\sigma_{2}$
has been set to zero.\cite{al203} For an insulator, the value of
$\sigma_{2}$ is negligible compared to $\sigma_{1}.$ The mass of
the core-shell structured particle has been computed as 
\begin{equation}
m=\underbrace{\frac{4}{3}\pi a^{3}\rho_{m,1}}_{m_{c}}+\underbrace{\frac{4}{3}\pi\left(b^{3}-a^{3}\right)\rho_{m,2}}_{m_{s}},\label{eq:mass-formula}
\end{equation}
 where $m_{c}$ and $m_{s}$ represent the masses of the core and
the shell, respectively. With these values assigned for each of the
parameters, Eq. (\ref{eq:ODE-rela-OR}) is solved via Runge-Kutta
method subjected to the following initial conditions, 
\begin{equation}
z_{d}\left(0\right)=0.25h\quad\textnormal{and}\quad\dot{z}_{d}\left(0\right)=0,\label{eq:initial-condition-zd}
\end{equation}
 which conditions are schematically illustrated in Fig. \ref{fig:initial_configuration}. 

In this work, it is assumed that core-shell structured charged-particle
is already ionized. And, for the initial conditions, this ionized
particle just happens to be at $z_{d}=0.25h$ with no initial speed.
Then, the question to be asked in this work is this; what happens
to the dynamics of this charged particle afterwards? Before I present
the result of dynamics, I shall discuss how such particle might be
ionized experimentally. 

Szirmai demonstrated that a core-shell structured particle, i.e.,
an aluminum core surrounded by a thin layer of aluminum oxide, can
be charged by exposing it to a static electric field of sufficiently
large strength. At the onset of the field emission, electrons are
stripped from the particle's conductive core and tunnel through a
thin dielectric layer, leaving the particle net positively charged
as a whole.\cite{STM-theory} Because electrons are more strongly
bonded in a dielectric, it takes significantly larger electric field
to ionize insulators. Therefore, it is reasonable to assume no free
surface charges on the insulating shell in Fig. \ref{fig:initial_configuration},
implying $\sigma_{2}=0\,\textnormal{C}\cdot\textnormal{m}^{-2}$ there. 

\begin{figure}[h]
\begin{centering}
\includegraphics[width=1\columnwidth]{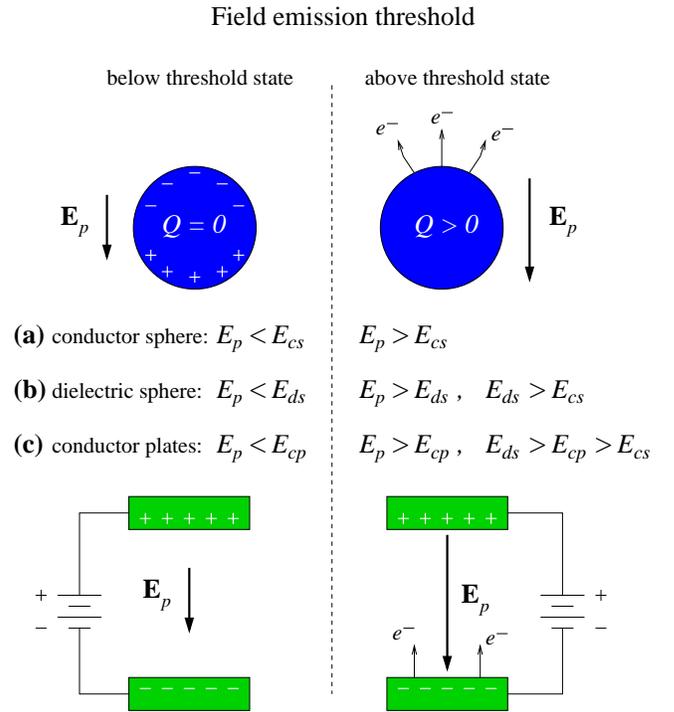}
\par\end{centering}

\caption{(Color online) Schematic of electron field emission threshold. When
the magnitude $E_{p}\equiv\left\Vert \mathbf{E}_{p}\right\Vert $
of applied electric field is less than the field emission threshold,
spherical particle is polarized inside, but no electrons are field
emitted and the particle as a whole remains neutral, i.e., $Q=0.$
When $E_{p}$ exceeds the field emission threshold, electrons are
stripped (or field emitted) from the particle, leaving the particle
net positively charged, i.e., $Q>0.$ Similarly, when $E_{p}$ is
below the field emission threshold, no electrons are emitted from
the lower potential plate. However, when $E_{p}$ exceeds the field
emission threshold, electrons from lower potential plate begin to
tunnel across the vacuum gap and electric discharge sets in. In the
figure, field emission electrons are indicated by $e^{-}.$ \label{fig:field-emission-process}}
\end{figure}

The field emission process strongly depends on the geometry as well
as on the orientation of host material.\cite{Raichev-FE,Jonge-FE,Jonge-2-FE,Li-Cheng-FE,Eda-FE}
For instance, a spherical surface has a lower field emission threshold
point than a flat surface; which is also the reason why a conductive
needle emits electron better than a thick rod conductor. Physically,
a sharp tip can be described by a surface with large curvature whereas
the dull one is described by a surface with smaller curvature. By
definition, the curvature of a circle of radius $r$ is large for
small $r$ and is small for large $r.$ Accordingly, spherical nanoparticles
have very large curvature whereas macroscopic spherical particles
have very small curvature. What is referred to as a flat surface is
just a special case in which the radius $r$ of a sphere becomes infinite
in extent. Since the conductive needle emits electrons better than
a thick rod conductor, a needle has a lower field emission threshold
than a rod. This implies that smaller spherical conductors, such as
nanoparticles, have lower field emission threshold than larger spherical
conductors or flat surfaces. That explained, the field emission thresholds
are schematically summarized in Fig. \ref{fig:field-emission-process}
for a spherical conductor, spherical dielectric, and a conductor plate,
wherein the following field emission thresholds are assumed: (a) $E_{cs}$
for the conducting sphere, (b) $E_{ds}$ for the dielectric sphere,
and (c) $E_{cp}$ for the conductor plate in vacuum. Since the geometry
of conductor plates is just an extension of infinitely large sphere,
the finite sized conducting spheres have lower field emission threshold
than large conductor plates; and, thus, $E_{cp}>E_{cs}.$ Because
electrons are very strongly bonded in a dielectric, exceptionally
large electric field must be applied to strip an electron from a dielectric
material. In general, it is much easier to strip an electron from
a conductor plate than from a dielectric. Hence, it is reasonable
to assume $E_{ds}>E_{cp}.$ In summary, the three field emission thresholds
satisfy the inequality given by 
\[
E_{ds}>E_{cp}>E_{cs}>0,
\]
 where $E_{ds},$ $E_{cp},$ and $E_{cs}$ are field emission thresholds
for dielectric sphere, conductor plate, and conductor sphere, respectively. 

What if the conducting spherical particle is coated with a thin layer
of a dielectric shell, just like the one illustrated in Fig. \ref{fig:initial_configuration};
would it be still possible to ionize such particle by a process of
field emission? The answer to this is yes, of course. For instance,
Konopsky et al. have experimentally measured field emissions from
sharp silicon tips covered with thin dielectric calcium fluoride layers.\cite{STM-2-Si-tip-coated-CaF2}
Iwasaki and Sudoh investigated electron tunneling through an aluminum
oxide thin film on a nickel-aluminum metal composite.\cite{STM-3-Al2O3-on-NiAl}
And, Kurnosikov et al. have also investigated electron transport through
alumina oxide tunnel junctions.\cite{STM-4-Al2O3-tunneling} Nevertheless,
the insulating dielectric shell on the surface of spherical conductor
increases the minimum electric field required to ionize a particle.
Then, to ionize a core-shell structured particle, it is crucial that
two plane-parallel conductor plates are able to produce electric field
large enough to ionize a particle without electric discharge setting
in. When there is a field emission originating from one of the conductor
plates, i.e., the lower conductor plate in Fig. \ref{fig:initial_configuration},
the positively ionized core-shell structured particle quickly neutralizes.
To prevent this, the two plane-parallel conductor plates must be able
to produce sufficiently large electric field to ionize the particle,
but at the same time this electric field must not be large enough
to initiate field emission from conductor plates themselves. 

Zouache and Lefort investigated the phenomenon of electric discharge
in plane-parallel conductor plates, such as the one illustrated in
Fig. \ref{fig:initial_configuration}, but without particle inside.\cite{IEEE-vacuum}
In their configuration, two plates are separated by an empty space
gap of one micron in length. They have tested various materials for
the conductor plates. Among various materials tried for conductor
plates, one was prepared from a mixture of $60\textnormal{\%}$ silver
and $40\textnormal{\%}$ nickel in its material composition. Conductor
plates with such material composition generated electric discharge
at applied electric field strength of $E_{p}\approx3850\,\textnormal{MV}\cdot\textnormal{m}^{-1}.$
In a vacuum gap separated plane-parallel conductor plates, electric
discharge is attributed to electrons from lower potential plate tunneling
through the gap towards the plate with higher potential, thereby shorting
the two plates and causing an electric discharge. In this regard,
the phenomenon of electric discharge is intrinsically connected to
the field emission threshold illustrated in Fig. \ref{fig:field-emission-process}(c).
Hence, in the aforementioned Zouache and Lefort's configuration, the
field emission threshold of $E_{cp}\lesssim3850\,\textnormal{MV}\cdot\textnormal{m}^{-1}$
can be roughly approximated. This field emission threshold value may
be compared with the electric field used by Szirmai to ionize his
particle. In Szirmai's experiment, an electric field of $E_{p}\approx4.67\,\textnormal{MV}\cdot\textnormal{m}^{-1}$
was sufficient to produce a net charge of $\sim4\times10^{-15}\,\textnormal{C}$
on a core-shell structured spherical aluminum particle of $3\,\mu\textnormal{m}$
in diameter. For the particle in Szirmai's experiment, the field emission
threshold of $E_{cs}\lesssim4.67\,\textnormal{MV}\cdot\textnormal{m}^{-1}$
can be assumed. Comparing the two values, one finds 
\[
\frac{E_{cp}}{E_{cs}}=\frac{3850\,\textnormal{MV}\cdot\textnormal{m}^{-1}}{4.67\,\textnormal{MV}\cdot\textnormal{m}^{-1}}\approx824,
\]
 which result suggests that for the configuration illustrated in Fig.
\ref{fig:initial_configuration}, the particle can be sufficiently
ionized when Zouache and Lefort's conductor plates are used. Further,
by limiting applied electric field to a value much smaller than $E_{cp}$
and, yet, much larger than $E_{cs},$ the particle can be ionized
in the absence of field emission electrons originating from the plate
held at lower potential. This prevents the ionized particle from neutralizing.
For instance, the electric field value of $E_{p}\approx0.5E_{cp}$
or $E_{p}\approx412E_{cs}$ is much smaller than $E_{cp},$ but it
is still much larger than $E_{cs}.$ 

\begin{figure}[h]
\begin{centering}
\includegraphics[width=1\columnwidth]{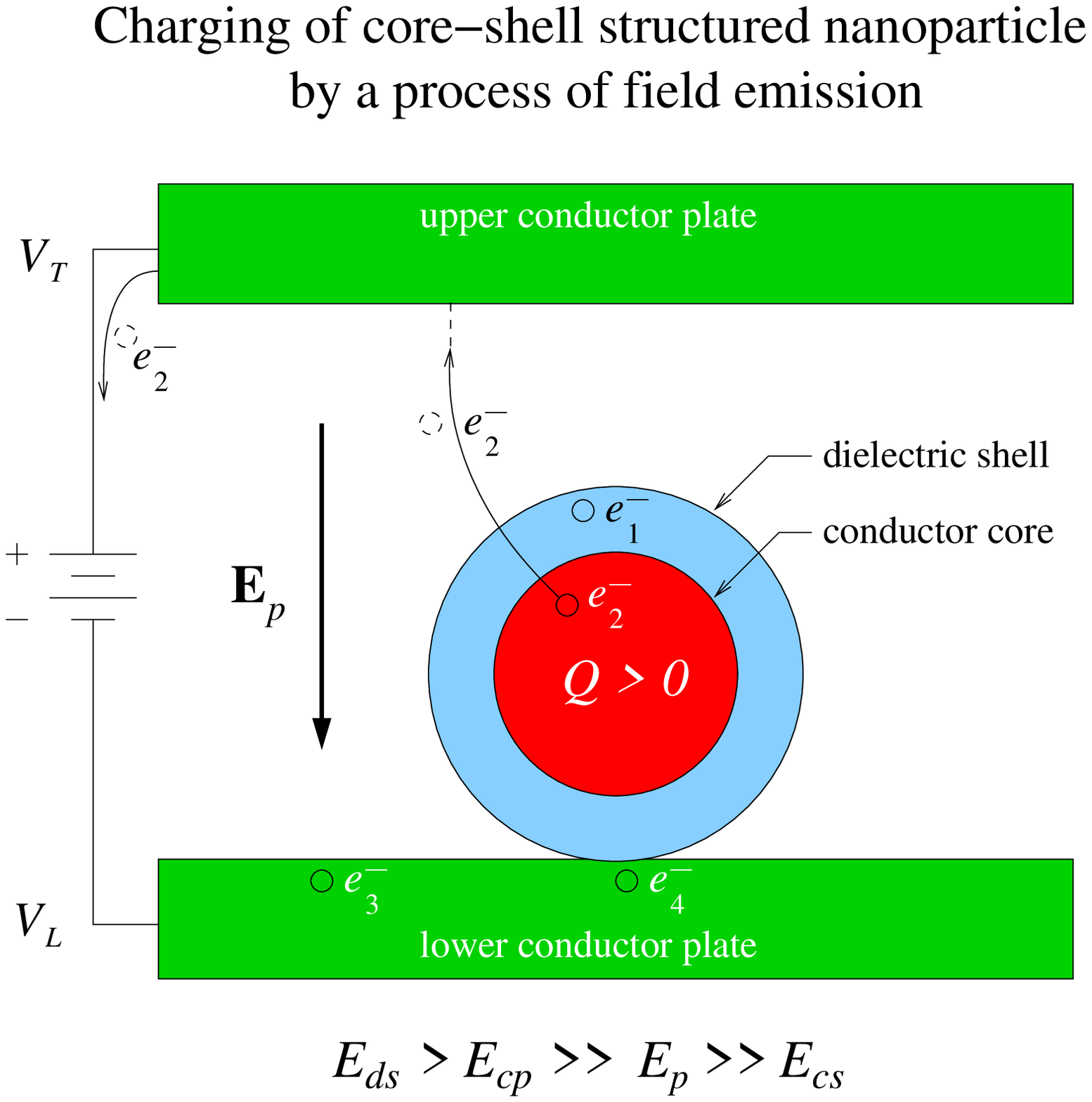}
\par\end{centering}

\caption{(Color online) Schematic illustration of ionization process of core-shell
structured nanoparticle. For the illustration purpose, particular
electrons in the particle's shell and the core regions are labeled
$e_{1}^{-}$ and $e_{2}^{-},$ respectively. Similarly, particular
electrons in the lower conductor plate are labeled $e_{3}^{-}$ and
$e_{4}^{-}.$ Initially, the particle as a whole is neutral and $Q=0.$
After the field emission, the particle as a whole becomes net positively
charged, $Q>0.$ \label{fig:ionization-process-of-CSCP}}
\end{figure}

Aforementioned process of charging by field emission is schematically
summarized in Fig. \ref{fig:ionization-process-of-CSCP}. To illustrate
the mechanism, particular electrons in the particle's shell and the
core regions are labeled $e_{1}^{-}$ and $e_{2}^{-},$ respectively.
Similarly, particular electrons in the lower conductor plate are labeled
$e_{3}^{-}$ and $e_{4}^{-}.$ It is understood that particle as a
whole is initially electrically neutral, implying $Q=0.$ That cleared,
an applied static electric field of magnitude $E_{p}\equiv\left\Vert \mathbf{E}_{p}\right\Vert ,$
where 
\begin{equation}
E_{p}\gg E_{cs},\;\textnormal{ but }\; E_{p}\ll E_{cp}<E_{ds},\label{eq:remain-positive-condition}
\end{equation}
 is produced by connecting two plane-parallel conductor plates to
a battery. Since $E_{p}\gg E_{cs},$ where $E_{cs}$ is the field
emission threshold for the particle's conductor core (see Fig. \ref{fig:field-emission-process}),
$e_{2}^{-}$ tunnels through the dielectric shell and the particle
as a whole becomes net positively charged. Because $E_{p}\ll E_{ds},$
where $E_{ds}$ is the field emission threshold for the particle's
dielectric shell, no electrons are physically stripped from the shell
and $e_{1}^{-}$ remains confined to the shell, leaving the shell
electrically neutral. Similarly, because $E_{p}\ll E_{cp},$ where
$E_{cp}$ is the field emission threshold for the lower conductor
plate, no electrons can escape the surface of the lower conductor
plate. The electrons, i.e., $e_{3}^{-}$ and $e_{4}^{-},$ may redistribute
themselves, but cannot physically escape the conductor plate's surface;
which is the reason why $e_{4}^{-}$ cannot tunnel through the particle's
dielectric shell to neutralize the positively charged core. To do
so, $e_{4}^{-}$ must first escape the plate's surface, which is not
possible since $E_{p}\ll E_{cp}.$ In conclusion, once the particle
is charged and the electric field satisfies the condition defined
in Eq. (\ref{eq:remain-positive-condition}), the core-shell structured
particle remains positively charged. 

Figure \ref{fig:ionization-process-of-CSCP} presents a way to prepare
yet another initial configuration, which is different from the one
illustrated in Fig. \ref{fig:initial_configuration}, for the differential
equation of Eq. (\ref{eq:ODE-rela-OR}). I shall briefly outline how
such initial configuration might be prepared from Fig. \ref{fig:ionization-process-of-CSCP}.
Starting with initially uncharged core-shell structured particle,
the particle is ionized following the scheme illustrated in Fig. \ref{fig:ionization-process-of-CSCP}.
Once the core-shell structured particle has been sufficiently ionized,
the entire system is physically flipped over. Don't worry about the
particle falling down because it won't. In fact, the positively charged
particle sticks to the surface of the conductor plate held at lower
of the two potentials. The reason for this is explained later in this
section when the types of forces involved in the system are discussed.
With the configuration illustrated in Fig. \ref{fig:ionization-process-of-CSCP}
flipped over, the initial conditions are now specified by 
\[
z_{d}\left(0\right)=0\quad\textnormal{and}\quad\dot{z}_{d}\left(0\right)=0.
\]
 In the flipped over configuration, the applied electric field $\mathbf{E}_{p}$
is in the $\mathbf{e}_{z}$ direction. Now, to make this configuration
suitable for the differential equation of Eq. (\ref{eq:ODE-rela-OR}),
the direction of applied electric field must be reversed. Thus, reversing
the direction of $\mathbf{E}_{p}$ to $-\mathbf{e}_{z}$ direction,
this configuration becomes identical to the initial configuration
illustrated in Fig. \ref{fig:initial_configuration}, except now $z_{d}\left(0\right)=0$
instead of $z_{d}\left(0\right)=0.25h.$ Either choice is good for
the initial condition of $z_{d}\left(t\right).$ That said, I shall
keep using the initial conditions specified in Eq. (\ref{eq:initial-condition-zd})
and the initial configuration illustrated in Fig. \ref{fig:initial_configuration}
for the rest of this work. 

\begin{figure}[h]
\begin{centering}
\includegraphics[width=1\columnwidth]{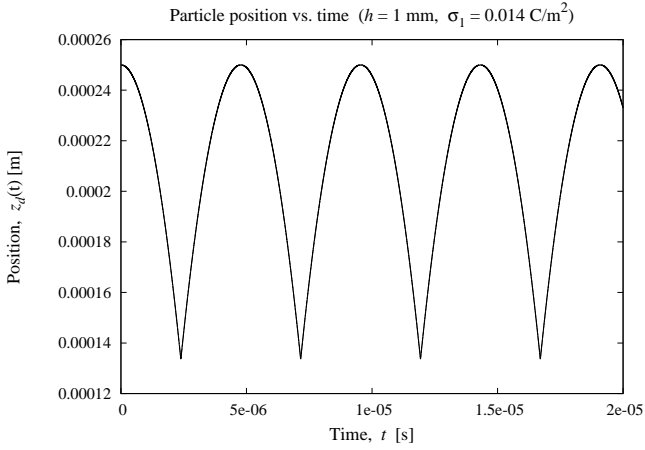}
\par\end{centering}

\caption{Particle distance from the surface of upper conductor plate as function
of time. For the plot, the values defined in Eq. (\ref{eq:parameter_VALUE})
have been used with initial conditions specified in Eq. (\ref{eq:initial-condition-zd}).
$V_{T}-V_{L}=8\,\textnormal{kV}$ implies an applied electric field
of $E_{p}=8\,\textnormal{MV}\cdot\textnormal{m}^{-1}.$ The upper
electrode is located at $z_{d}=0\,\textnormal{m}.$ \label{fig:zd(t)-plot}}
\end{figure}

Returning to the plotting of $z_{d}$ from Eq. (\ref{eq:ODE-rela-OR})
with initial conditions specified in Eq. (\ref{eq:initial-condition-zd}),
the particle position as function of time has been plotted in Fig.
\ref{fig:zd(t)-plot} using the parameter values defined in Eq. (\ref{eq:parameter_VALUE}).
The upper electrode is located at $z_{d}=0\,\textnormal{m}$ in the
plot. As it can be observed from the plot, the core-shell structured
charged-particle executes an oscillatory motion between the two plane-parallel
electrodes; and, such motion does not involve charge exchanges. One
knows this because the governing equation of motion, i.e., Eq. (\ref{eq:ODE-rela-OR}),
has been derived without any assumption of charge exchange. Such oscillatory
behavior is fundamentally different from the traditional picture,
which process assumes charge exchange mechanisms. The core-shell structured
charged-particle rebounds at $z_{d}\approx2.5\times10^{-4}\,\textnormal{m}$
from the upper conductor plate. This rebounding position is too far
from the lower conductor plate to account for any charge exchange
processes even in the physical situations. In the plot of Fig. \ref{fig:zd(t)-plot},
the lower conductor plate is located at $z_{d}\approx h=0.001\,\textnormal{m}.$ 

Hereafter, it is understood that charged-particle oscillation does
not involve any charge exchanges for the rest of this paper. That
said, what's the criterion for charged-particle oscillation? Oscillatory
modes occur when the effective charge carried by the positively charged
particle satisfies the condition given by 
\[
Q_{T}>\frac{2\pi\epsilon_{0}\kappa_{3}}{\xi}\left(\left|\psi\right|E_{p}+\sqrt{\psi^{2}E_{p}^{2}+\frac{16mg\xi}{\pi\epsilon_{0}\kappa_{3}}}\right),
\]
 where 
\[
\xi=\frac{1}{z_{d,m}^{2}}-\frac{1}{\left(h-z_{d,m}\right)^{2}}>0,
\]
 
\[
\psi=\frac{\gamma\left(b^{3}-a^{3}\right)-b^{3}}{z_{d,m}^{3}}+\frac{\gamma\left(b^{3}-a^{3}\right)-b^{3}}{\left(h-z_{d,m}\right)^{3}}-8<0,
\]
 and 
\[
\gamma=\frac{3\kappa_{3}b^{3}}{\left(\kappa_{2}+2\kappa_{3}\right)b^{3}+2\left(\kappa_{2}-\kappa_{3}\right)a^{3}}<1.
\]
 Similarly, for a negatively charged core-shell structured particle,
the oscillatory criterion is given by 
\[
\left|Q_{T}\right|>\frac{2\pi\epsilon_{0}\kappa_{3}}{\eta}\left(\left|\psi\right|E_{p}-\sqrt{\psi^{2}E_{p}^{2}-\frac{16mg\eta}{\pi\epsilon_{0}\kappa_{3}}}\right),
\]
 where 
\[
\eta=\frac{1}{\left(h-z_{d,m}\right)^{2}}-\frac{1}{z_{d,m}^{2}}>0.
\]

The mass dependence in Eqs. (\ref{eq:OSC-CRITERION}) and (\ref{eq:OSC-CRITERION-nC})
reveals that particle with larger mass requires significantly larger
effective charge compared to the particle with smaller mass to initiate
oscillatory motion. Initially neutral particle can be charged or ionized
by exposing it to a strong static electric field. For a nanoparticle,
large portion of atoms composing it participate in the ionization,
yielding in relatively large charge density per mass. However, for
a macroscopic particle or an object, a great portion of atoms composing
it does not participate in the ionization process and only those near
the surface participate in the ionization due to electric field shielding
effects. As a result, macroscopic particles have relatively small
charge density per mass. One may argue that the strength of electric
field can always be increased to completely ionize the macroscopic
particle. That, however, is not possible because, even in vacuum,
electric breakdown sets in at some point and everything neutralizes.\cite{IEEE-vacuum} 

Based on this argument, the oscillation criterion specified in Eq.
(\ref{eq:OSC-CRITERION}) is more likely to be satisfied by microscopic
or smaller particles than by macroscopic counterparts. This implies
the charged-particle oscillation presented in this paper is more likely
to be observed from nanoparticle systems than from systems involving
macroscopic particles. I shall now discuss why microscopic particles
are more likely to satisfy the criterion of Eqs. (\ref{eq:OSC-CRITERION})
or (\ref{eq:OSC-CRITERION-nC}) than the macroscopic counterparts.
To demonstrate this, I shall consider an aluminum ball of radius $b=1.5\,\mu\textnormal{m}$
representing the smaller particle and another one with radius $b=3\,\textnormal{mm}$
representing the larger counterpart. To keep matters simple, I shall
assume that the space between electrodes is a vacuum. 

In vacuum, $\kappa_{3}=1$ and $\pi\epsilon_{0}\kappa_{3}=2.78\times10^{-11}\,\textnormal{N}^{-1}\cdot\textnormal{m}^{-2}\cdot\textnormal{C}^{2},$
and the gravity constant is $g=9.8\,\textnormal{m}\cdot\textnormal{s}^{-2}.$
The mass densities of aluminum and aluminum oxide are $\rho_{m,1}=2700\,\textnormal{kg}\cdot\textnormal{m}^{-3}$
and $\rho_{m,2}=3800\,\textnormal{kg}\cdot\textnormal{m}^{-3},$ respectively.
Thus, for a core radius of $a=1.5\,\mu\textnormal{m}$ and shell thickness
of $b-a=4\,\textnormal{nm},$ the total mass of the particle is obtained
using Eq. (\ref{eq:mass-formula}) to yield $m\approx3.86\times10^{-14}\,\textnormal{kg}.$
The dielectric constant for the particle's shell is $\kappa_{2}=6.$
For the value of $z_{d,m},$ I shall choose $z_{d,m}=0.25h,$ where
$h=1\,\textnormal{mm}.$ For the applied electric field, I shall choose
$E_{p}=8\,\textnormal{kV}\cdot\textnormal{m}^{-1}.$ Insertion of
these values into Eq. (\ref{eq:OSC-CRITERION}) yields 
\begin{align}
Q_{T,mic}\equiv Q_{T} & >5\times10^{-13}\,\textnormal{C},\label{eq:QT_micro}
\end{align}
 where the notation $Q_{T,mic}$ denotes the microscopic particle. 

Now, I shall compute the same for the macroscopic counterpart. An
aluminum ball of core radius $a=3\,\textnormal{mm}$ and shell thickness
of $b-a=4\,\textnormal{nm}$ has a mass of $m\approx3.05\times10^{-4}\,\textnormal{kg},$
where Eq. (\ref{eq:mass-formula}) has been used to compute the mass.
To make sure this aluminum ball has sufficient room between the electrodes
for oscillation, the gap between the two electrodes is increased to
a value of $h=1\,\textnormal{m}.$ Keeping all other values same as
previous, Eq. (\ref{eq:OSC-CRITERION}) gives 
\begin{align}
Q_{T,mac}\equiv Q_{T} & >9\times10^{-7}\,\textnormal{C},\label{eq:QT_macro}
\end{align}
 where the notation $Q_{T,mac}$ denotes the macroscopic particle.
Is this an experimentally obtainable value? The answer to this is
maybe. It depends on what kind of electrodes are being used. Even
in vacuum, one cannot increase the strength of electric field indefinitely
without electrical breakdown setting in, beyond which point everything
neutralizes.\cite{IEEE-vacuum} Comparing the two results, Eqs. (\ref{eq:QT_micro})
and (\ref{eq:QT_macro}), $Q_{T,mac}$ is greater than $Q_{T,mic}$
by factor of a million. This result alone shows that the kind of charged-particle
oscillation mechanism presented here, i.e., one that does not involve
charge transfer processes, is most likely to be observed from microscopic
or smaller particles than from macroscopic counterparts. 

\begin{figure}[h]
\begin{centering}
\includegraphics[width=1\columnwidth]{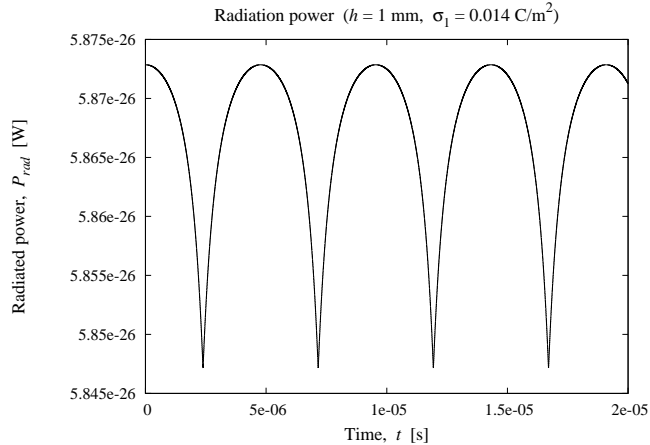}
\par\end{centering}

\caption{The profile of Liénard radiation power corresponding to the oscillating
charged-particle illustrated in Fig. \ref{fig:zd(t)-plot}. \label{fig:Lienard-radiation-power}}
\end{figure}

Because the system illustrated in Fig. \ref{fig:particle-in-capacitor}(a)
involves charged-particle executing an oscillatory motion, it radiates
electromagnetic energy; and, the power of radiated energy can be obtained
from Liénard radiation formula, 
\begin{align*}
P_{rad} & =\frac{8\pi\epsilon_{0}\kappa_{3}^{2}\nu^{2}}{3c^{3}}\left(1-\frac{\dot{z}_{d}^{2}}{c^{2}}\right)^{-3}\ddot{z}_{d}^{2}.
\end{align*}
 With the explicit expression for $\ddot{z}_{d}$ inserted from Eq.
(\ref{eq:ODE-rela-OR}), this becomes 

\begin{align}
P_{rad} & =\frac{8\pi\epsilon_{0}\kappa_{3}^{2}\nu^{2}}{3c^{3}}\left(\frac{\pi\epsilon_{0}\kappa_{3}\nu}{4m}\left\{ \frac{\nu}{\left(z_{d}+b\right)^{2}}\right.\right.\nonumber \\
 & -\frac{\nu}{\left(h-z_{d}-b\right)^{2}}+\frac{\left[\gamma\left(b^{3}-a^{3}\right)-b^{3}\right]E_{p}}{\left(z_{d}+b\right)^{3}}\nonumber \\
 & \left.\left.+\frac{\left[\gamma\left(b^{3}-a^{3}\right)-b^{3}\right]E_{p}}{\left(h-z_{d}-b\right)^{3}}-8E_{p}\right\} -g\right)^{2}.\label{eq:radiation-power}
\end{align}
 The profile of Liénard radiation power corresponding to the oscillating
core-shell structured charged-particle illustrated in Fig. \ref{fig:zd(t)-plot}
has been computed using Eq. (\ref{eq:radiation-power}). The result
shows train of emitted radiation power as the particle oscillates,
as illustrated in Fig. \ref{fig:Lienard-radiation-power}. 

It is worthwhile to discuss the types of forces involved in the plot
of Fig. \ref{fig:zd(t)-plot}. The force responsible for generating
such particle motion is Eq. (\ref{eq:FT-OR-with-QT}) (or Eq. (\ref{eq:FT-OR})),
of course. Since the force contribution from gravity is negligible
in the oscillation regime, I shall only work with $\mathbf{F}_{1}$
and $\mathbf{F}_{2}$ of Eqs. (\ref{eq:F1-with-QT}) and (\ref{eq:F2-with-QT}),
respectively, for the discussion here. The force $\mathbf{F}_{1}$
is composed of the following three force contributions, 
\begin{align*}
\mathbf{F}_{1} & =\mathbf{f}_{1,1}+\mathbf{f}_{1,2}+\mathbf{f}_{1,3},
\end{align*}
 where 
\begin{align*}
\mathbf{f}_{1,1} & =\mathbf{e}_{z}\frac{1}{16}\frac{Q_{T}\nu}{s^{2}},\\
\mathbf{f}_{1,2} & =\mathbf{e}_{z}\frac{1}{16}\frac{Q_{T}\left[\gamma\left(b^{3}-a^{3}\right)-b^{3}\right]}{s^{3}}E_{p},\\
\mathbf{f}_{1,3} & =-\mathbf{e}_{z}\frac{Q_{T}}{4}E_{p}.
\end{align*}
 From Eqs. (\ref{eq:gama-OR}) and (\ref{eq:nu-OR}), it can be seen
that 
\[
\nu>0\;\textnormal{ and }\;1>\gamma>0.
\]
 This implies, $\gamma\left(b^{3}-a^{3}\right)-b^{3}<0$ and the previous
forces can be expressed as 
\begin{align*}
\mathbf{f}_{1,1} & \sim\mathbf{e}_{z}\frac{1}{s^{2}},\\
\mathbf{f}_{1,2} & \sim-\mathbf{e}_{z}\frac{E_{p}}{s^{3}},\\
\mathbf{f}_{1,3} & \sim-\mathbf{e}_{z}E_{p}.
\end{align*}
 Physically, $\mathbf{f}_{1,1}$ represents the force between the
charged particle and the image charge formed on the surface of conductor
plate. Since the image (or induced) charge has opposite polarity,
this force attracts the charged particle towards the plate. And, since
$\mathbf{f}_{1,1}$ is in the direction of $\mathbf{e}_{z},$ this
confirms such action. The $\mathbf{f}_{1,3}$ physically represents
the force by electric field on the charged particle. Such force always
pushes a positively charged particle in the direction of electric
field. And, since $\mathbf{f}_{1,3}$ is in the direction of $-\mathbf{e}_{z},$
which is the direction of electric field, this also confirms such
property. The remaining force term, $\mathbf{f}_{1,2},$ is a direct
consequence of having a core-shell structured charged-particle. To
be more accurate, $\mathbf{f}_{1,2}$ is a consequence of having particle
with structure, which is not a ``point'' particle. Particles with
structure can be polarized by applied electric field and such property
gives rise to $\mathbf{f}_{1,2}.$ Consequently, this force vanishes
in the absence of applied electric field. For a positively charged
particle, this force is induced in the same direction as the applied
electric field. 

Similarly, the force $\mathbf{F}_{2}$ of Eq. (\ref{eq:F2-with-QT})
can be decomposed into the following three force contributions: 
\begin{align*}
\mathbf{F}_{2} & =\mathbf{f}_{2,1}+\mathbf{f}_{2,2}+\mathbf{f}_{2,3},
\end{align*}
 where 
\begin{align*}
\mathbf{f}_{2,1} & =-\mathbf{e}_{z}\frac{1}{16}\frac{Q_{T}\nu}{\left(h-s\right)^{2}},\\
\mathbf{f}_{2,2} & =\mathbf{e}_{z}\frac{1}{16}\frac{Q_{T}\left[\gamma\left(b^{3}-a^{3}\right)-b^{3}\right]}{\left(h-s\right)^{3}}E_{p},\\
\mathbf{f}_{2,3} & =-\mathbf{e}_{z}\frac{Q_{T}}{4}E_{p}.
\end{align*}
 Since $\gamma\left(b^{3}-a^{3}\right)-b^{3}<0,$ these can be expressed
as 
\begin{align*}
\mathbf{f}_{2,1} & \sim-\mathbf{e}_{z}\frac{1}{s^{2}},\\
\mathbf{f}_{2,2} & \sim-\mathbf{e}_{z}\frac{E_{p}}{s^{3}},\\
\mathbf{f}_{2,3} & \sim-\mathbf{e}_{z}E_{p}.
\end{align*}
 Physically, $\mathbf{f}_{2,1}$ is the force between the image charge
and the charged particle. Since $\mathbf{f}_{2,1}$ is the force arising
when charged particle is close to the lower conductor plate, this
force must be directed towards the lower conductor plate. Since $\mathbf{f}_{2,1}$
is directed in $-\mathbf{e}_{z},$ this confirms such requirement.
The $\mathbf{f}_{2,3}$ is the force on the positively charged particle
due to the presence of electric field. Since electric field is in
the $-\mathbf{e}_{z}$ direction, so is $\mathbf{f}_{2,3},$ as it
must. Lastly, $\mathbf{f}_{2,2}$ is a consequence of having a charged-particle
with structure and not a point particle. Particles with structure
can be polarized by applied electric field and such property gives
rise to $\mathbf{f}_{2,2}.$ Again, this force is always in the direction
of applied electric field for a positively charged particle. 

\begin{figure}[h]
\begin{centering}
\includegraphics[width=1\columnwidth]{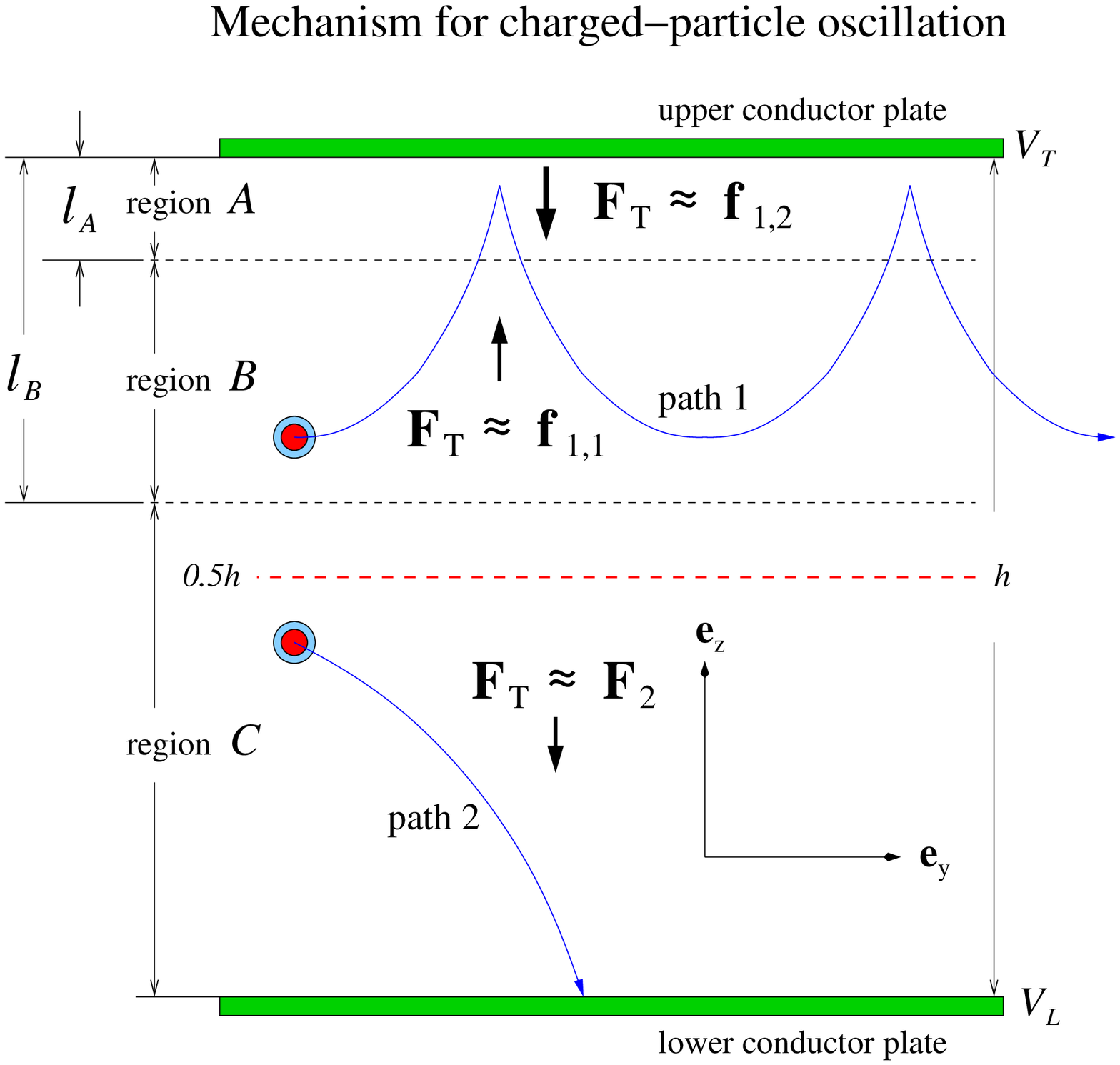}
\par\end{centering}

\caption{(Color online) Different forces dominate in each regions. In region
$A,$ the dominant force is $\mathbf{f}_{1,2}.$ In region $B,$ the
dominant force is $\mathbf{f}_{1,1}.$ In region $C,$ the dominant
force is $\mathbf{F}_{2}.$ The $\textnormal{path 1}$ and $\textnormal{path 2}$
represent the plots of $z_{d}\left(t\right)$ versus time graph, where
the time parameter is the horizontal axis. Here, $V_{T}>V_{L}.$ \label{fig:mechanism}}
\end{figure}

So, what is responsible for charged-particle oscillation? The second
force, i.e., $\mathbf{F}_{2}$ of Eq. (\ref{eq:F2-with-QT}), cannot
be responsible for charged-particle oscillation because $\mathbf{f}_{2,1},$
$\mathbf{f}_{2,2},$ and $\mathbf{f}_{2,3}$ are all directed in the
$-\mathbf{e}_{z}$ direction. However, the first force, i.e., $\mathbf{F}_{1}$
of Eq. (\ref{eq:F1-with-QT}), can give rise to charged-particle oscillatory
modes. This is because force $\mathbf{F}_{1}$ contains $\mathbf{f}_{1,1},$
which force points in the opposite direction of $\mathbf{f}_{1,2}$
and $\mathbf{f}_{1,3}.$  It is this competition between $\mathbf{f}_{1,1}$
and the other forces, i.e., $\mathbf{f}_{1,2}$ and $\mathbf{f}_{1,3},$
that puts core-shell structured charged-particle in an oscillatory
motion. Such mechanism is schematically illustrated in Fig. \ref{fig:mechanism}.
In region $A,$ the dominant force is $\mathbf{f}_{1,2}$ and the
magnitude of this force falls off like $\sim1/s^{3}$ with distance,
$s.$ However, in region $B,$ contributions from $\mathbf{f}_{1,2}$
weakens rapidly with distance and the force is dominated by $\mathbf{f}_{1,1},$
which force's magnitude falls off like $\sim1/s^{2}$ with distance.
Then, assuming the positively charged core-shell structured particle
satisfies the oscillation criterion specified in Eq. (\ref{eq:OSC-CRITERION}),
the particle cannot escape region $B$ and enter region $C,$ which
region contains no oscillation modes and any particles initially in
such region ends up sticking to the lower conductor plate, as schematically
illustrated in Fig. \ref{fig:mechanism}. 

That said, a positively charged core-shell structured particle initially
in region $B$ (or region $A$) would oscillate; and, the trace of
such oscillatory motion over time would be represented by the $\textnormal{path 1}$
illustrated in Fig. \ref{fig:mechanism}. Here, the $\textnormal{path 1}$
represents the plot of $z_{d}\left(t\right)$ versus time graph, where
the time parameter is the horizontal axis. On the other hand, the
same particle initially in region $C$ would not have any oscillations,
but it would follow the trace of the $\textnormal{path 2},$ where
the $\textnormal{path 2}$ represents the plot of $z_{d}\left(t\right)$
versus time graph with horizontal axis being the time axis. 

The sharp cusp in region $A$ can be explained from the fact that
$\mathbf{f}_{1,2}\sim-\left(1/s^{3}\right)\mathbf{e}_{z}$ is an extremely
short range force. Since the magnitude of such  force goes like $\sim1/s^{3},$
it can generate very large impulse over short time. Since $\mathbf{f}_{1,2}$
is in $-\mathbf{e}_{z}$ direction, the particle is repulsed from
the surface with very large force occurring over very short period.
However, in region $B,$ this force decays extremely rapidly and $\mathbf{f}_{1,1}\sim\left(1/s^{2}\right)\mathbf{e}_{z}$
dominates there. The direction of $\mathbf{f}_{1,1}$ is in $\mathbf{e}_{z};$
hence, the particle is pulled back to the upper conductor plate. This
process repeats itself, resulting in an oscillatory motion. 

\begin{figure}[h]
\begin{centering}
\includegraphics[width=1\columnwidth]{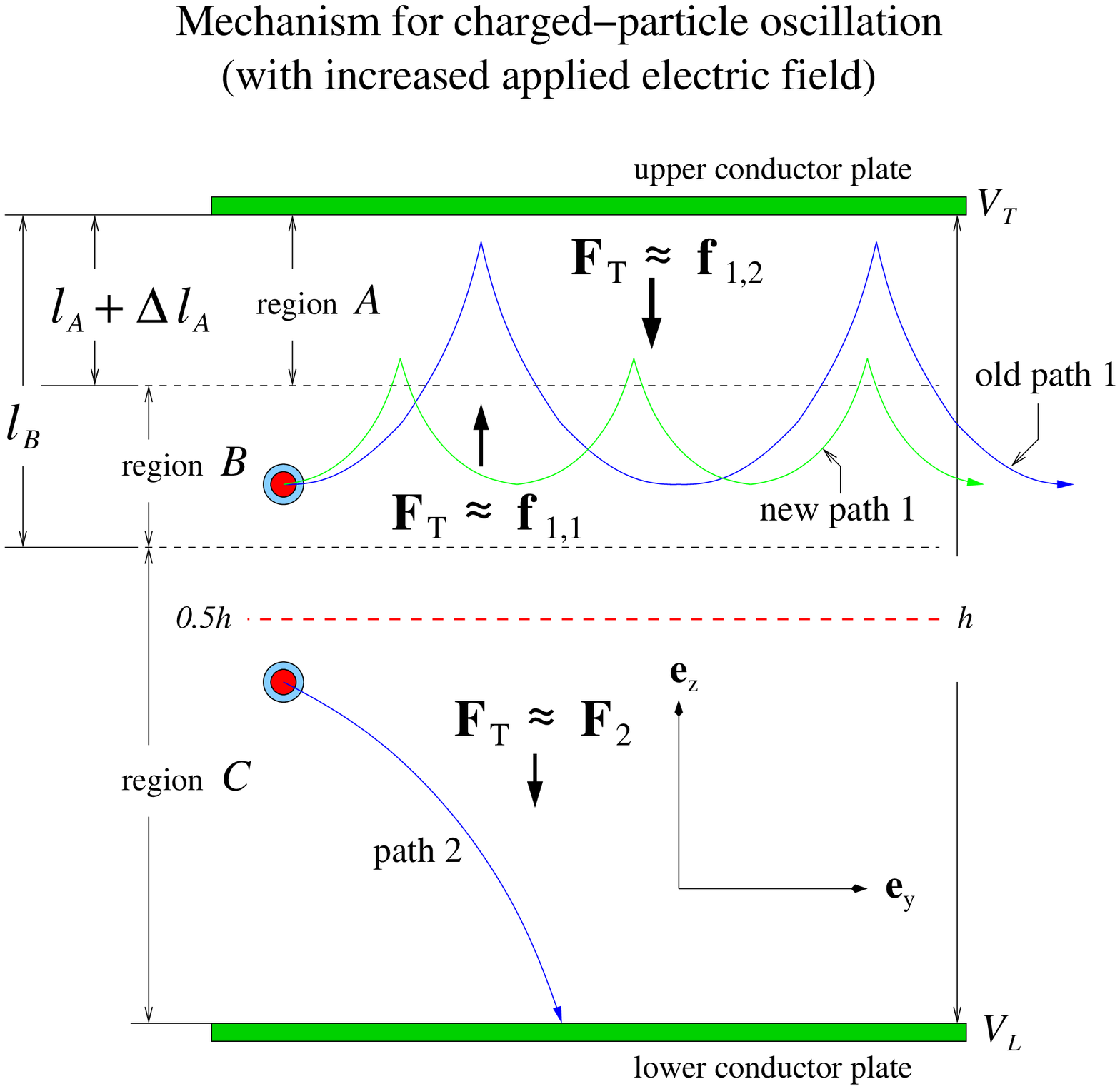}
\par\end{centering}

\caption{(Color online) The width of region $A$ has been increased to $l_{A}+\Delta l_{A}$
as a result of increased strength of applied electric field, $\mathbf{E}_{p}=-\mathbf{e}_{z}E_{p}.$
The $\textnormal{new path 1,}$ $\textnormal{old path 1,}$ and $\textnormal{path 2}$
represent the plots of $z_{d}\left(t\right)$ versus time graph, where
the time parameter is the horizontal axis. Here, $V_{T}>V_{L}.$ \label{fig:mechanism-2}}
\end{figure}

\begin{figure}[h]
\begin{centering}
\includegraphics[width=1\columnwidth]{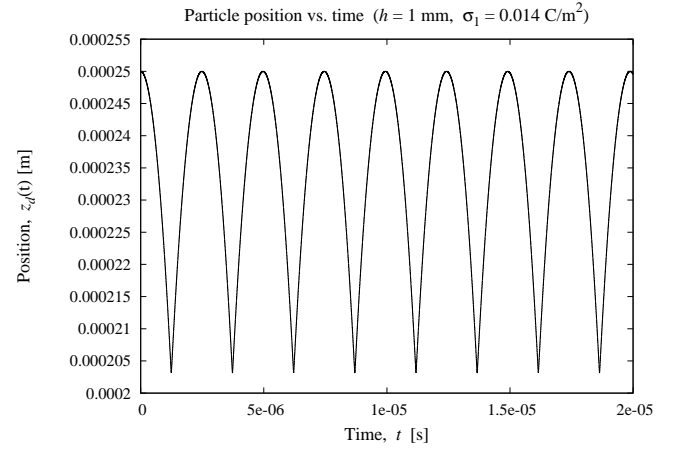}
\par\end{centering}

\caption{Particle distance from the surface of upper conductor plate as function
of time. For the plot, the values defined in Eq. (\ref{eq:parameter_VALUE})
have been used. Here, however, the value of $V_{T}-V_{L}$ has been
increased to $V_{T}-V_{L}=12\,\textnormal{kV},$ implying an applied
electric field strength of $E_{p}=12\,\textnormal{MV}\cdot\textnormal{m}^{-1}.$
The upper electrode is located at $z_{d}=0\,\textnormal{m}.$ \label{fig:zd(t)-plot-df}}
\end{figure}

What happens when the magnitude of applied electric field, $E_{p},$
is increased? The force $\mathbf{f}_{1,1}\sim\left(1/s^{2}\right)\mathbf{e}_{z}$
is independent of $E_{p};$ therefore, $l_{B}$ does not change in
Fig. \ref{fig:mechanism}. On the other hand, the force $\mathbf{f}_{1,2}\sim-\left(1/s^{3}\right)\mathbf{e}_{z}$
has an explicit dependence on the applied electric field, i.e., 
\[
\mathbf{f}_{1,2}\sim-\mathbf{e}_{z}\frac{E_{p}}{s^{3}};
\]
 therefore, the $l_{A}$ in Fig. \ref{fig:mechanism} gets increased
to $l_{A}+\Delta l_{A}$ with increased $E_{p}.$ Here, $l_{A}$ and
$l_{B}$ represent the locations of borderlines for regions $A$ and
$B,$ respectively. The result is that charged-particle trapped inside
the region $B$ is now forced to rebound more frequently due to the
fact that the width of region $B$ has been decreased by $\Delta l_{A}.$
Consequently, the frequency of charged-particle oscillation increases
with increased applied electric field; and, this is schematically
illustrated in Fig. \ref{fig:mechanism-2}, where it shows that $\textnormal{new path 1}$
has higher frequency than the $\textnormal{old path 1}.$ To validate
this, Eq. (\ref{eq:ODE-rela-OR}) has been plotted using the identical
settings used to obtain the result in Fig. \ref{fig:zd(t)-plot}.
This time, however, the strength of applied electric field has been
increased from $8\,\textnormal{MV}\cdot\textnormal{m}^{-1}$ to $12\,\textnormal{MV}\cdot\textnormal{m}^{-1}.$
The result is shown in Fig. \ref{fig:zd(t)-plot-df}. This result
can be compared with the one in Fig. \ref{fig:zd(t)-plot}, which
was obtained for $E_{p}=8\,\textnormal{MV}\cdot\textnormal{m}^{-1}.$
For the case of $E_{p}=12\,\textnormal{MV}\cdot\textnormal{m}^{-1},$
the frequency of charged-particle oscillation has been approximately
doubled compared to that of the case of $E_{p}=8\,\textnormal{MV}\cdot\textnormal{m}^{-1}.$
Notice that although the oscillation frequency has been approximately
doubled, its amplitude has been nearly halved. This must be so because
the particle's oscillation frequency had been increased as a result
of reduced width of region $B$ (or increased width of region $A$). 

Now, one cannot increase $E_{p}$ indefinitely to obtain higher oscillation
frequencies because, eventually, the width of region $B$ would become
zero. And, beyond that point, the charged-particle enters the region
$C$ and ends up sticking to the surface of the lower conductor plate.
In region $C,$ there are no oscillatory modes so any charged particle
in that region gets attracted to the surface of lower conductor plate
and stays there indefinitely. 

\begin{figure}[h]
\begin{centering}
\includegraphics[width=1\columnwidth]{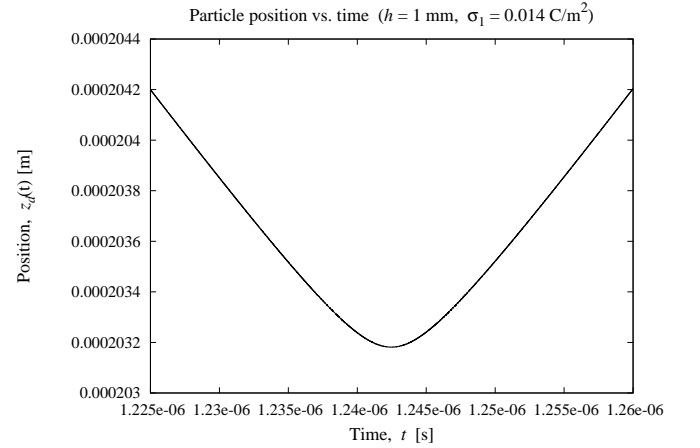}
\par\end{centering}

\caption{Particle distance from the surface of upper conductor plate as function
of time. The first sharp turning point near $z_{d}\approx0.000205\,\textnormal{m}$
in Fig. \ref{fig:zd(t)-plot-df} has been enlarged for a view, which
shows a smoothly varying curve. \label{fig:zd(t)-plot-df-zoom}}
\end{figure}

Having explained the kind of forces involved in the charged-particle
oscillation, i.e., plots generated in Figs. (\ref{fig:zd(t)-plot})
and (\ref{fig:zd(t)-plot-df}), it is now clear why there are cusps
in the plot. The shape of these sharp turning points can be deceiving
because these points are not really what they appear to be. In fact,
these turning points are smoothly varying points and this is illustrated
in Fig. \ref{fig:zd(t)-plot-df-zoom}, where one of such sharp points
has been enlarged for a view. At these points, the magnitude of force
experienced by the particle falls off with distance like $\sim1/s^{3},$
where $s$ is the distance between the particle's center of mass and
the rebounding plate's surface. At very short separation distances,
this repulsion force becomes extremely impulsive over very short period.
But, nevertheless, $\sim1/s^{3}$ is still a well behaved function
because $s$ cannot become zero, as the particle cannot touch the
surface of rebounding conductor plate. Doing so would require an infinite
energy, which is not possible. 

\begin{figure}[h]
\begin{centering}
\includegraphics[width=1\columnwidth]{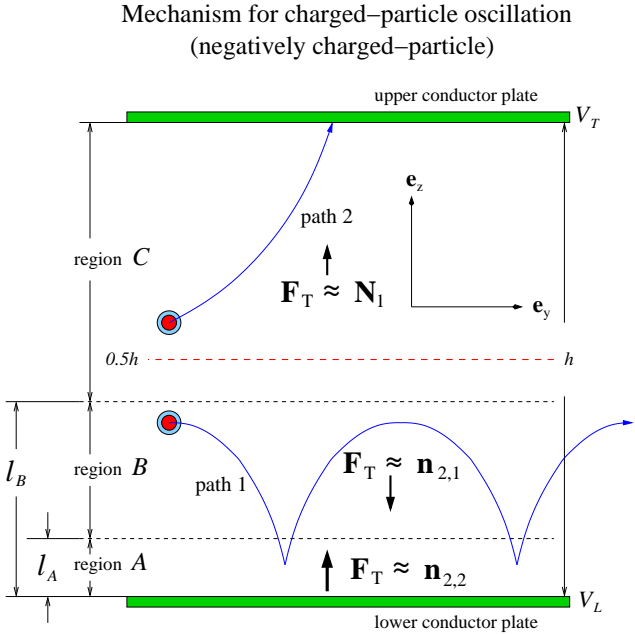}
\par\end{centering}

\caption{(Color online) For the case of negatively charged-particle, the oscillation
modes exist near the lower conductor plate, which is exactly opposite
of the positively charged-particle case (see Fig. \ref{fig:mechanism}).
In region $A,$ the dominant force is $\mathbf{n}_{2,2};$ and, in
region $B,$ the dominant force is $\mathbf{n}_{2,1}.$ The $\textnormal{path 1}$
and $\textnormal{path 2}$ represent the plots of $z_{d}\left(t\right)$
versus time graph, where the time parameter is the horizontal axis.
Here, $V_{T}>V_{L}.$ \label{fig:mechanism-nC}}
\end{figure}

What happens when the core-shell structured particle is negatively
charged? In that case, the $\mathbf{F}_{1}$ and $\mathbf{F}_{2}$
of Eqs. (\ref{eq:F1-with-QT}) and (\ref{eq:F2-with-QT}) get modified
as 
\begin{align*}
\mathbf{F}_{1} & =\mathbf{e}_{z}\frac{\left|Q_{T}\right|}{4}\left[\frac{\left|\nu\right|}{4s^{2}}+\frac{\left|\gamma\left(b^{3}-a^{3}\right)-b^{3}\right|E_{p}}{4s^{3}}+E_{p}\right]
\end{align*}
 and 
\begin{align*}
\mathbf{F}_{2} & =\mathbf{e}_{z}\frac{\left|Q_{T}\right|}{4}\left[\frac{\left|\gamma\left(b^{3}-a^{3}\right)-b^{3}\right|E_{p}}{4\left(h-s\right)^{3}}-\frac{\left|\nu\right|}{4\left(h-s\right)^{2}}+E_{p}\right].
\end{align*}
 To distinguish the analysis here from the previous case involving
a positively charged core-shell structured particle, I shall rewrite
$\mathbf{F}_{1}$ and $\mathbf{F}_{2}$ as 
\begin{align*}
\mathbf{N}_{1} & =\mathbf{e}_{z}\frac{\left|Q_{T}\right|}{4}\left[\frac{\left|\nu\right|}{4s^{2}}+\frac{\left|\gamma\left(b^{3}-a^{3}\right)-b^{3}\right|E_{p}}{4s^{3}}+E_{p}\right]
\end{align*}
 and 
\begin{align*}
\mathbf{N}_{2} & =\mathbf{n}_{2,1}+\mathbf{n}_{2,2}+\mathbf{n}_{2,3},
\end{align*}
 where 
\[
\mathbf{n}_{2,1}=-\mathbf{e}_{z}\frac{\left|Q_{T}\right|}{16}\frac{\left|\nu\right|}{\left(h-s\right)^{2}}\sim-\mathbf{e}_{z}\frac{1}{\left(h-s\right)^{2}},
\]
 
\[
\mathbf{n}_{2,2}=\mathbf{e}_{z}\frac{\left|Q_{T}\right|}{16}\frac{\left|\gamma\left(b^{3}-a^{3}\right)-b^{3}\right|}{\left(h-s\right)^{3}}E_{p}\sim\mathbf{e}_{z}\frac{E_{p}}{\left(h-s\right)^{3}},
\]
 
\[
\mathbf{n}_{2,3}=\mathbf{e}_{z}\frac{\left|Q_{T}\right|}{4}E_{p}\sim\mathbf{e}_{z}E_{p}.
\]
 Since all of the terms are positive in force $\mathbf{N}_{1},$ it
cannot generate any oscillations. On the other hand, the force $\mathbf{N}_{2}$
contains both positive and negative terms; and, therefore, it can
generate oscillatory modes. Such is schematically illustrated in Fig.
\ref{fig:mechanism-nC}. 

\begin{figure}[h]
\begin{centering}
\includegraphics[width=1\columnwidth]{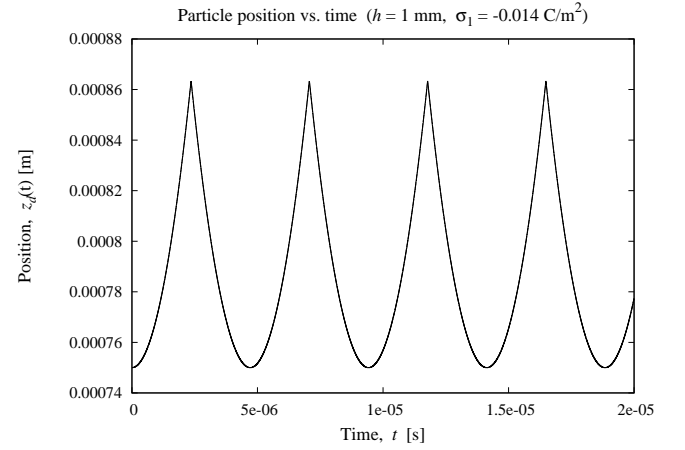}
\par\end{centering}

\caption{Particle distance from the surface of upper conductor plate as function
of time. The lower electrode is located at $z_{d}=0.001\,\textnormal{m}$
and the upper electrode is located at $z_{d}=0\,\textnormal{m}.$
The charged particle is negatively charged and it is oscillating near
the lower electrode. \label{fig:zd(t)-plot-nC}}
\end{figure}

\begin{figure}[h]
\begin{centering}
\includegraphics[width=1\columnwidth]{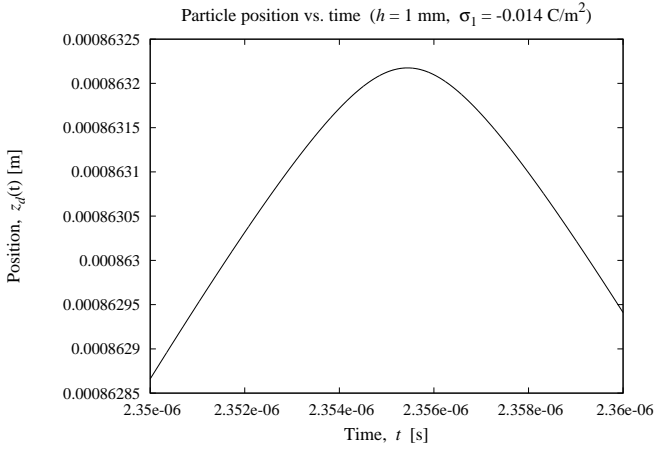}
\par\end{centering}

\caption{Particle distance from the surface of upper conductor plate as function
of time. The first sharp turning point near $z_{d}\approx0.00086\,\textnormal{m}$
in Fig. \ref{fig:zd(t)-plot-nC} has been enlarged for a view, which
shows a smoothly varying curve. \label{fig:zd(t)-plot-nC-zoom}}
\end{figure}

To validate the argument illustrated in Fig. \ref{fig:mechanism-nC},
Eq. (\ref{eq:ODE-rela-OR}) has been plotted using the same values
specified in Eq. (\ref{eq:parameter_VALUE}), except now $\sigma_{1}=-0.014\,\textnormal{C}\cdot\textnormal{m}^{-2}.$
Also, to account for the oscillatory motion, the initial conditions
have been specified as 
\[
z_{d}\left(0\right)=0.75h\quad\textnormal{and}\quad\dot{z}_{d}\left(0\right)=0.
\]
Such initial conditions have been chosen because oscillatory modes
only exist for $z_{d}>0.5h$ for negatively charged particles. The
result is plotted in Fig. \ref{fig:zd(t)-plot-nC}, where it shows
negatively charged core-shell structured particle oscillating near
the lower conductor plate. In the plot, the lower conductor plate
is located at $z_{d}=0.001\,\textnormal{m}$ and the upper conductor
plate is located at $z_{d}=0\,\textnormal{m}.$ To show that charged-particle
trajectory is represented by a well behaved function, one of the sharp
turning points in Fig. \ref{fig:zd(t)-plot-nC} has been enlarged
for inspection. As it can be seen in Fig. \ref{fig:zd(t)-plot-nC-zoom},
 the cusp looking points are deceiving because these are smoothly
varying points. The magnitude of force acting on the particle near
these points falls off with distance like $\sim1/\left(h-s\right)^{3}.$
When the negatively charged core-shell structured particle is very
close to the lower conductor plate, $h-s$ becomes very small and
this results in a very large force that acts to repulse the particle
from the surface of the lower conductor plate. Nonetheless, this is
a well defined force because $h-s$ cannot become zero. Doing so would
require an infinite energy, which is not possible. 

This briefly summarizes the essence of this investigation. To complete
the task, I shall now work out the detailed derivations of key solutions
used in this article. I shall begin by solving the boundary value
problem for the electrostatic potentials in regions $M_{1},$ $M_{2},$
and $M_{3}$ of Fig. \ref{fig:particle-in-capacitor}.

\section{Theory}

\subsection{Free charge distribution }

\begin{figure}[h]
\begin{centering}
\includegraphics[width=1\columnwidth]{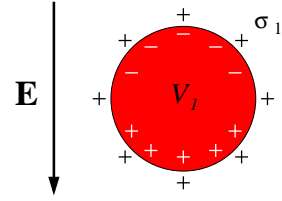}
\par\end{centering}

\caption{(Color online) Charge distributions of (a) electrically neutral spherical
conductor and (b) a positively charged spherical conductor with surface
free charge density, $\sigma_{1},$ in an otherwise uniform electric
field, $\mathbf{E}_{p}.$ \label{fig:charge-distribution}}
\end{figure}

The correct specification of electric charge distribution is of crucial
importance in any electrostatic boundary value problem. When an uncharged,
electrically neutral, spherical conductor is placed in an otherwise
uniform electric field, the charges inside the conductor redistribute
such that the potential $V_{1}$ is a constant there, as illustrated
in Fig. \ref{fig:charge-distribution}(a). By definition, in an electrically
neutral conductor, every charges are paired with one with opposite
polarity. Therefore, the spherical conductor illustrated in Fig. \ref{fig:charge-distribution}(a),
as a whole, is electrically neutral. 

Now, how do charges get distributed when you place a positively charged
spherical conductor in an otherwise uniform, constant, electric field?
By definition, an electrically charged conductor has excess number
of charges of one polarity that cannot be paired with one with opposite
polarity. The paired ones do whatever they can to make net electric
field zero inside the conductor. The result is that the paired ones
redistribute as illustrated in Fig. \ref{fig:charge-distribution}(a).
What about the excess, unpaired, charges of same polarity? These must
be redistributed such that the potential $V_{1}$ is a constant inside
the conductor. One such distribution, perhaps the only one, is illustrated
in Fig. \ref{fig:charge-distribution}(b). Assuming the charged spherical
conductor has only the surface ``free charges'' and no volume ``free
charges,'' i.e., no excess charges embedded inside the volume, the
free charges on the surface of spherical conductor must be uniformly
distributed over the entire spherical surface else the net electric
field inside the spherical conductor would not be a zero. As an alternate
explanation, the spherical conductor in Fig. \ref{fig:charge-distribution}(a),
including its surface, represents an equipotential surface. When an
excess of free charges of same polarity, say positive charges, is
placed on such an equipotential surface, the charges get instantaneously
redistributed over the surface due to Coulomb repulsion between the
charges. The result is that these charges are uniformly distributed
over the equipotential surface, as illustrated in Fig. \ref{fig:charge-distribution}(b). 

That explained, I shall assume that surface ``free charge'' density
$\sigma_{1}$ in Fig. \ref{fig:particle-in-capacitor} is a constant
which is uniformly distributed over the surface $r=a$ of the spherical
conductor core throughout this work. To generalize the problem, the
dielectric shell surrounding the spherical core in Fig. \ref{fig:particle-in-capacitor}
is allowed for a surface ``free charge'' density $\sigma_{2}$ at
$r=b.$ Introduction of free charge on the surface of dielectric shell
is purely academic. For realistic dielectrics, $\sigma_{2}$ is negligibly
small, if not zero. Hence, it can always be set to zero in the final
solution. Therefore, I shall keep the problem simple by assuming that
$\sigma_{2}$ is a constant which is also uniformly distributed over
the surface $r=b$ of the dielectric shell throughout this work. 

Lastly, although the illustration in Fig. \ref{fig:charge-distribution}
considered only an excess positive charge case in which $Q_{T}>0,$
the treatment throughout this work is not limited to such case only.
The effective charge $Q_{T}$ can have either positive or negative
polarities.

\subsection{Derivation of electrostatic potentials}

The apparatus for the problem is illustrated in Fig. \ref{fig:particle-in-capacitor}(a),
where a core-shell structured charged-particle is placed between two
DC voltage biased plane-parallel conductors. Electrostatic potentials
in regions $M_{1},$ $M_{2},$ and $M_{3}$ are described by Laplace
equation, 
\[
\nabla^{2}V=0.
\]
 In spherical polar coordinate system, Fig. \ref{fig:particle-in-capacitor}(b),
Laplace equation reads 
\begin{align*}
\frac{1}{r^{2}}\frac{\partial}{\partial r}\left(r^{2}\frac{\partial V}{\partial r}\right)+\frac{1}{r^{2}\sin\theta}\frac{\partial}{\partial\theta}\left(\sin\theta\frac{\partial V}{\partial\theta}\right)\\
+\frac{1}{r^{2}\sin^{2}\theta}\frac{\partial^{2}V}{\partial\phi^{2}} & =0.
\end{align*}
 For the system with azimuth symmetry, 
\begin{align*}
\frac{\partial V}{\partial\phi} & =0,
\end{align*}
 the Laplace equation reduces to 
\begin{equation}
\frac{\partial}{\partial r}\left(r^{2}\frac{\partial V}{\partial r}\right)+\frac{1}{\sin\theta}\frac{\partial}{\partial\theta}\left(\sin\theta\frac{\partial V}{\partial\theta}\right)=0.\label{eq:Laplace-PDE}
\end{equation}
 Equation (\ref{eq:Laplace-PDE}) has the general solution given by
\[
V\left(r,\theta\right)=\sum_{\ell=0}^{\infty}\left(A_{\ell}r^{\ell}+\frac{B_{\ell}}{r^{\ell+1}}\right)P_{\ell},
\]
 where coefficients $A_{\ell}$ and $B_{\ell}$ are constants, and
$P_{\ell}\equiv P_{\ell}\left(\cos\theta\right)$ is the Legendre
polynomial of order $\ell.$ For regions $M_{1},$ $M_{2},$ and $M_{3}$
in Fig. \ref{fig:particle-in-capacitor}(a), the electrostatic potentials
are given by 
\begin{align}
V_{1}\left(r,\theta\right) & =\sum_{\ell=0}^{\infty}A_{\ell}r^{\ell}P_{\ell},\label{eq:V1-pre}\\
V_{2}\left(r,\theta\right) & =\sum_{\ell=0}^{\infty}\left(B_{\ell}r^{\ell}+\frac{C_{\ell}}{r^{\ell+1}}\right)P_{\ell},\label{eq:V2-pre}\\
V_{3}\left(r,\theta\right) & =\sum_{\ell=0}^{\infty}\left(D_{\ell}r^{\ell}+\frac{E_{\ell}}{r^{\ell+1}}\right)P_{\ell},\label{eq:V3-pre}
\end{align}
 where coefficients $A_{\ell},$ $B_{\ell},$ $C_{\ell},$ $D_{\ell},$
and $E_{\ell}$ are to be joined together by appropriate boundary
conditions at the interfaces between regions. Equation (\ref{eq:V1-pre})
does not contain terms like $\sim r^{-\ell-1}$ because these terms
blow up at the origin. 

Electrostatic potential inside of a conductor is constant; and, therefore,
Eq. (\ref{eq:V1-pre}) becomes 
\begin{equation}
V_{1}\left(r,\theta\right)=A_{0}\equiv V_{0},\label{eq:V1-10}
\end{equation}
 where $V_{0}$ is a constant. Potential is continuous at $r=a,$
\[
V_{2}\left(a,\theta\right)=V_{1}\left(a,\theta\right).
\]
 From Eqs. (\ref{eq:V2-pre}) and (\ref{eq:V1-10}), it can be shown
that 
\[
B_{0}+\frac{C_{0}}{a}+\sum_{\ell=1}^{\infty}\left(B_{\ell}a^{\ell}+\frac{C_{\ell}}{a^{\ell+1}}\right)P_{\ell}=V_{0}
\]
 or 
\[
\left(B_{0}-V_{0}+\frac{C_{0}}{a}\right)P_{0}+\sum_{\ell=1}^{\infty}\left(B_{\ell}a^{\ell}+\frac{C_{\ell}}{a^{\ell+1}}\right)P_{\ell}=0,
\]
 where $P_{0}=1.$ It follows that each Legendre polynomials ($P_{0},$
$P_{1},$ $P_{2},$ and so on) are linearly independent functions;
and, therefore, the coefficient of each Legendre polynomials must
be zero independently else this algebraic equation cannot be satisfied.
Thus, 
\begin{align*}
B_{0}+C_{0}a^{-1} & =V_{0},\\
B_{\ell}a^{\ell}+C_{\ell}a^{-\ell-1} & =0.
\end{align*}
 Solving for $C_{0}$ and $C_{\ell}$ yields 
\begin{align*}
C_{0} & =a\left(V_{0}-B_{0}\right),\\
C_{\ell} & =-B_{\ell}a^{2\ell+1},\quad\ell\geq1.
\end{align*}
 From these results, Eq. (\ref{eq:V2-pre}) becomes 
\begin{align}
V_{2}\left(r,\theta\right) & =B_{0}\left(1-\frac{a}{r}\right)+\frac{aV_{0}}{r}\nonumber \\
 & +\sum_{\ell=1}^{\infty}B_{\ell}\left(r^{\ell}-\frac{a^{2\ell+1}}{r^{\ell+1}}\right)P_{\ell}.\label{eq:V2-10}
\end{align}
 Equations (\ref{eq:V3-pre}) and (\ref{eq:V2-10}) must be continuous
at $r=b,$ 
\[
V_{2}\left(b,\theta\right)=V_{3}\left(b,\theta\right).
\]
 With Eqs. (\ref{eq:V3-pre}) and (\ref{eq:V2-10}), it can be shown
that 
\begin{align*}
B_{0}\left(1-\frac{a}{b}\right)+\frac{aV_{0}}{b}+\sum_{\ell=1}^{\infty}B_{\ell}\left(b^{\ell}-\frac{a^{2\ell+1}}{b^{\ell+1}}\right)P_{\ell}\\
=D_{0}+\frac{E_{0}}{b}+\sum_{\ell=1}^{\infty}\left(D_{\ell}b^{\ell}+\frac{E_{\ell}}{b^{\ell+1}}\right)P_{\ell}
\end{align*}
 or 
\begin{align*}
0 & =\left[B_{0}\left(1-\frac{a}{b}\right)+\frac{aV_{0}}{b}-D_{0}-\frac{E_{0}}{b}\right]P_{0}\\
 & +\sum_{\ell=1}^{\infty}\left[B_{\ell}\left(b^{\ell}-\frac{a^{2\ell+1}}{b^{\ell+1}}\right)-D_{\ell}b^{\ell}-\frac{E_{\ell}}{b^{\ell+1}}\right]P_{\ell},
\end{align*}
 where $P_{0}=1.$ Again, Legendre polynomials ($P_{0},$ $P_{1},$
$P_{2},$ and so on) are linearly independent functions; and, therefore,
the coefficient of each Legendre polynomials must vanish independently
else this algebraic equation cannot be satisfied. Hence, 
\begin{align*}
D_{0}+\frac{E_{0}}{b} & =B_{0}\left(1-\frac{a}{b}\right)+\frac{aV_{0}}{b},\\
D_{\ell}b^{\ell}+\frac{E_{\ell}}{b^{\ell+1}} & =B_{\ell}\left(b^{\ell}-\frac{a^{2\ell+1}}{b^{\ell+1}}\right),
\end{align*}
 and the following coefficients are obtained: 
\begin{align*}
E_{0} & =B_{0}\left(b-a\right)+aV_{0}-D_{0}b,\\
E_{\ell} & =B_{\ell}\left(b^{2\ell+1}-a^{2\ell+1}\right)-D_{\ell}b^{2\ell+1},\quad\ell\geq1.
\end{align*}
 Using these results, Eq. (\ref{eq:V3-pre}) becomes 
\begin{align}
V_{3}\left(r,\theta\right) & =D_{0}\left(1-\frac{b}{r}\right)+\frac{B_{0}\left(b-a\right)+aV_{0}}{r}\nonumber \\
 & +\sum_{\ell=1}^{\infty}\left[\vphantom{\frac{\frac{1}{1}}{\frac{1^{1}}{1^{2}}}}D_{\ell}\left(r^{\ell}-\frac{b^{2\ell+1}}{r^{\ell+1}}\right)\right.\nonumber \\
 & \left.+\frac{B_{\ell}\left(b^{2\ell+1}-a^{2\ell+1}\right)}{r^{\ell+1}}\vphantom{\frac{\frac{1}{1}}{\frac{1^{1}}{1^{2}}}}\right]P_{\ell}.\label{eq:V3-10}
\end{align}
Equation (\ref{eq:V3-10}) must simultaneously satisfy the boundary
conditions at the surfaces of the upper and the lower conductor plates
illustrated in Fig. \ref{fig:particle-in-capacitor}(a). 

In Cartesian coordinates, the surface of the upper conductor plate
is described by the $z=s$ plane and the surface of the lower conductor
plate is described by the $z=s-h$ plane. At distances sufficiently
far from the particle (or very close to the surface of conductor plates),
the potential inside of the parallel plates can be approximated as
\[
V_{p}=-\int_{s-h}^{z}\mathbf{E}_{p}\cdot\mathbf{e}_{z}dz^{\prime}+V_{L},
\]
 where $V_{L}$ is the voltage applied to the lower conductor plate,
$\mathbf{e}_{z}$ is the versor along the Cartesian $z$ axis, and
$\mathbf{E}_{p}$ is the electric field inside of the parallel  plates
in the absence of the charged-particle. The expression for $\mathbf{E}_{p}$
is given by 
\begin{equation}
\mathbf{E}_{p}=-\mathbf{e}_{z}\frac{1}{h}\left(V_{T}-V_{L}\right),\label{eq:Ep}
\end{equation}
 from which the $V_{p}$ can be obtained: 
\begin{equation}
V_{p}\left(z\right)=E_{p}\left(z-s+h\right)+V_{L},\label{eq:Vp-cartesian}
\end{equation}
 where 
\begin{equation}
E_{p}\equiv\left\Vert \mathbf{E}_{p}\right\Vert =\frac{1}{h}\left(\left|V_{T}-V_{L}\right|\right).\label{eq:Ep-magnitude}
\end{equation}
 In spherical polar coordinate system, the Cartesian coordinate $z$
is represented by 
\begin{align*}
z & =r\cos\theta
\end{align*}
 and Eq. (\ref{eq:Vp-cartesian}) becomes 
\begin{equation}
V_{p}\left(r,\theta\right)=E_{p}\left(r\cos\theta-s+h\right)+V_{L}.\label{eq:Vp-SPC}
\end{equation}
 For $r$ very large, but not infinite in extent, the contributions
from terms like $\sim r^{-1}$ and $\sim r^{-\ell-1}$ become negligible
in Eq. (\ref{eq:V3-10}) and the $V_{3}$ takes the form given by
\begin{equation}
V_{3}\left(r,\theta\right)\approx D_{0}+\sum_{\ell=1}^{\infty}D_{\ell}r^{\ell}P_{\ell},\label{eq:V3-at-very-large-r}
\end{equation}
 where $b\ll r<\infty.$ At distances sufficiently far from the particle,
$V_{3}\left(r,\theta\right)\approx V_{p}\left(r,\theta\right).$ 

One may recall a typical problem in electrodynamics, wherein a charged
sphere is immersed in an otherwise constant and uniform electric field.
For instance, assuming uniform electric field is along the $\mathbf{e}_{z}$
axis, a useful boundary condition is that at infinity, electric field
is just $\mathbf{E}=\mathbf{e}_{z}\left\Vert \mathbf{E}\right\Vert r\cos\theta.$
Now, at distances which are infinitesimally close to the conductor
plate's surface, electric field must be perpendicular to the plate's
surface. This is because the surface of conductor plate is an equipotential
surface and electric fields are perpendicular to the equipotential
surface by definition, of course. In this regard, Eq. (\ref{eq:Vp-SPC})
mimics the electric field boundary condition at infinity for the textbook
problem in electrodynamics, wherein a charged sphere is immersed in
an otherwise constant and uniform electric field and the student is
asked to solve for the potential around the sphere. 

That said, Eqs. (\ref{eq:Vp-SPC}) and (\ref{eq:V3-at-very-large-r})
are equated to yield 
\begin{align*}
 & D_{0}+D_{1}r\cos\theta+\sum_{\ell=2}^{\infty}D_{\ell}r^{\ell}P_{\ell}\\
 & \qquad\approx E_{p}r\cos\theta+E_{p}\left(h-s\right)+V_{L}.
\end{align*}
Matching the coefficients of the like Legendre polynomials yield 
\begin{align*}
D_{0} & \approx E_{p}\left(h-s\right)+V_{L},\\
D_{1} & \approx E_{p},\\
D_{\ell} & \approx0,\quad\ell\geq2.
\end{align*}
 Using these results, Eq. (\ref{eq:V3-10}) becomes 
\begin{align}
V_{3}\left(r,\theta\right) & \approx E_{p}\left(h-s\right)+V_{L}+E_{p}r\cos\theta\nonumber \\
 & +\left[B_{0}\left(b-a\right)+aV_{0}-bE_{p}\left(h-s\right)-bV_{L}\right]\frac{1}{r}\nonumber \\
 & +\left[B_{1}\left(b^{3}-a^{3}\right)-b^{3}E_{p}\right]\frac{\cos\theta}{r^{2}}\nonumber \\
 & +\sum_{\ell=2}^{\infty}\frac{B_{\ell}\left(b^{2\ell+1}-a^{2\ell+1}\right)}{r^{\ell+1}}P_{\ell},\label{eq:V3-20}
\end{align}
 where it is understood that $b\ll r<\infty.$ The electrostatic potential,
which satisfies the Laplace equation, is a second order differential
equation. Therefore, its derivatives must be satisfied at the boundaries.
The remaining unknowns, $B_{0},$ $B_{1},$ $B_{\ell}$ for $\ell\geq2,$
and $V_{0}$ are evaluated from the statement about the discontinuity
of electric displacement at $r=b$ and at $r=a.$ 

At $r=b,$ the normal component of the electric displacement suffers
a discontinuity given by  
\begin{equation}
\left.\left[\mathbf{e}_{r}\cdot\mathbf{D}_{3}\left(r,\theta\right)-\mathbf{e}_{r}\cdot\mathbf{D}_{2}\left(r,\theta\right)\right]\right|_{r=b}=\sigma_{2},\label{eq:continuity-of-D-at-r=00003Db}
\end{equation}
 where $\sigma_{2}$ is the surface free-charge density at $r=b,$
the $\mathbf{e}_{r}$ is a unit vector pointing in the radially outward
direction, and $\mathbf{D}_{2}$ and $\mathbf{D}_{3}$ represent electric
displacements in regions $M_{2}$ and $M_{3},$ respectively. In the
linear dielectric approximation, the electric displacement can be
expressed as 
\begin{equation}
\mathbf{D}_{i}\left(r,\theta\right)=-\epsilon_{0}\kappa_{i}\nabla V_{i}\left(r,\theta\right),\label{eq:electric-displacement-definition}
\end{equation}
 where $\kappa_{i}$ is the dielectric constant in region $M_{i}$
and $\epsilon_{0}$ is the electric permittivity of the free space.
Hence, Eq. (\ref{eq:continuity-of-D-at-r=00003Db}) can be expressed
as 
\begin{equation}
\left.\left[\kappa_{2}\mathbf{e}_{r}\cdot\nabla V_{2}\left(r,\theta\right)-\kappa_{3}\mathbf{e}_{r}\cdot\nabla V_{3}\left(r,\theta\right)\right]\right|_{r=b}=\frac{\sigma_{2}}{\epsilon_{0}}.\label{eq:continuity-of-D-at-r=00003Db-10}
\end{equation}
 In spherical polar coordinate system, the $\nabla$ operator is defined
by 
\[
\nabla=\mathbf{e}_{r}\frac{\partial}{\partial r}+\mathbf{e}_{\theta}\frac{1}{r}\frac{\partial}{\partial\theta}+\mathbf{e}_{\phi}\frac{1}{r\sin\theta}\frac{\partial}{\partial\phi}
\]
 and Eq. (\ref{eq:continuity-of-D-at-r=00003Db-10}) becomes 
\begin{equation}
\left.\left[\kappa_{2}\frac{\partial V_{2}\left(r,\theta\right)}{\partial r}-\kappa_{3}\frac{\partial V_{3}\left(r,\theta\right)}{\partial r}\right]\right|_{r=b}=\frac{\sigma_{2}}{\epsilon_{0}},\label{eq:Neumann-BC-r=00003Db}
\end{equation}
 which constitutes the Neumann boundary condition at $r=b.$ In explicit
forms, the derivatives in Eq. (\ref{eq:Neumann-BC-r=00003Db}) are
evaluated as 
\begin{align}
 & \frac{\partial V_{2}\left(r,\theta\right)}{\partial r}=\left(B_{0}-V_{0}\right)\frac{a}{r^{2}}\nonumber \\
 & +\sum_{\ell=1}^{\infty}B_{\ell}\left[\ell r^{\ell-1}+\frac{\left(\ell+1\right)a^{2\ell+1}}{r^{\ell+2}}\right]P_{\ell}\label{eq:derivative-V2}
\end{align}
 and 
\begin{align}
 & \frac{\partial V_{3}\left(r,\theta\right)}{\partial r}=E_{p}\cos\theta\nonumber \\
 & -\left[B_{0}\left(b-a\right)+aV_{0}-bE_{p}\left(h-s\right)-bV_{L}\right]\frac{1}{r^{2}}\nonumber \\
 & -2\left[B_{1}\left(b^{3}-a^{3}\right)-b^{3}E_{p}\right]\frac{\cos\theta}{r^{3}}\nonumber \\
 & -\sum_{\ell=2}^{\infty}\frac{\left(\ell+1\right)B_{\ell}\left(b^{2\ell+1}-a^{2\ell+1}\right)}{r^{\ell+2}}P_{\ell},\label{eq:derivative-V3}
\end{align}
 where Eqs. (\ref{eq:V2-10}) and (\ref{eq:V3-20}) have been used.
Insertion of Eqs. (\ref{eq:derivative-V2}) and (\ref{eq:derivative-V3})
into Eq. (\ref{eq:Neumann-BC-r=00003Db}) yields 
\begin{align*}
\frac{\sigma_{2}}{\epsilon_{0}} & =\frac{1}{b^{2}}\left\{ B_{0}\left[a\left(\kappa_{2}-\kappa_{3}\right)+b\kappa_{3}\right]\right.\\
 & \left.-a\left(\kappa_{2}-\kappa_{3}\right)V_{0}-b\kappa_{3}E_{p}\left(h-s\right)-b\kappa_{3}V_{L}\right\} \\
 & +\kappa_{2}\left(1+\frac{2a^{3}}{b^{3}}\right)B_{1}\cos\theta\\
 & +\kappa_{3}\left\{ \frac{2}{b^{3}}\left[B_{1}\left(b^{3}-a^{3}\right)-b^{3}E_{p}\right]-E_{p}\right\} \cos\theta\\
 & +\sum_{\ell=2}^{\infty}B_{\ell}P_{\ell}\left\{ \frac{\kappa_{2}\ell}{b^{1-\ell}}\right.\\
 & \left.+\frac{\ell+1}{b^{\ell+2}}\left[a^{2\ell+1}\left(\kappa_{2}-\kappa_{3}\right)+b^{2\ell+1}\kappa_{3}\right]\right\} 
\end{align*}
 or 
\begin{align*}
0 & =\left(\frac{1}{b^{2}}\left\{ B_{0}\left[a\left(\kappa_{2}-\kappa_{3}\right)+b\kappa_{3}\right]-b\kappa_{3}V_{L}\right.\right.\\
 & \left.\left.-a\left(\kappa_{2}-\kappa_{3}\right)V_{0}-b\kappa_{3}E_{p}\left(h-s\right)\right\} -\frac{\sigma_{2}}{\epsilon_{0}}\right)P_{0}\\
 & +\kappa_{2}\left(1+\frac{2a^{3}}{b^{3}}\right)B_{1}P_{1}\\
 & +\kappa_{3}\left\{ \frac{2}{b^{3}}\left[B_{1}\left(b^{3}-a^{3}\right)-b^{3}E_{p}\right]-E_{p}\right\} P_{1}\\
 & +\sum_{\ell=2}^{\infty}B_{\ell}P_{\ell}\left\{ \frac{\kappa_{2}\ell}{b^{1-\ell}}\right.\\
 & \left.+\frac{\ell+1}{b^{\ell+2}}\left[a^{2\ell+1}\left(\kappa_{2}-\kappa_{3}\right)+b^{2\ell+1}\kappa_{3}\right]\right\} ,
\end{align*}
 where $P_{0}=1$ and $P_{1}=\cos\theta.$ Because each Legendre polynomials
of order $\ell$ are linearly independent functions, this algebraic
relation can be satisfied if and only if the coefficients of each
Legendre  polynomials vanish independently. Hence, 
\begin{align*}
B_{0}\left[a\left(\kappa_{2}-\kappa_{3}\right)+b\kappa_{3}\right]-a\left(\kappa_{2}-\kappa_{3}\right)V_{0}\\
-b\kappa_{3}\left[E_{p}\left(h-s\right)+V_{L}\right]-\frac{b^{2}\sigma_{2}}{\epsilon_{0}} & =0,
\end{align*}
 
\begin{align*}
\kappa_{3}\left\{ \frac{2}{b^{3}}\left[B_{1}\left(b^{3}-a^{3}\right)-b^{3}E_{p}\right]-E_{p}\right\} \\
+\kappa_{2}\left(1+\frac{2a^{3}}{b^{3}}\right)B_{1} & =0,
\end{align*}
 and 
\begin{align*}
B_{\ell}\left\{ \frac{\ell+1}{b^{\ell+2}}\left[a^{2\ell+1}\left(\kappa_{2}-\kappa_{3}\right)+b^{2\ell+1}\kappa_{3}\right]+\frac{\kappa_{2}\ell}{b^{1-\ell}}\right\}  & =0.
\end{align*}
 One reads off immediately that 
\begin{align}
B_{0} & =\frac{b\kappa_{3}\left[E_{p}\left(h-s\right)+V_{L}\right]+b^{2}\epsilon_{0}^{-1}\sigma_{2}}{a\left(\kappa_{2}-\kappa_{3}\right)+b\kappa_{3}}\nonumber \\
 & +\frac{a\left(\kappa_{2}-\kappa_{3}\right)V_{0}}{a\left(\kappa_{2}-\kappa_{3}\right)+b\kappa_{3}},\label{eq:B0-pre}
\end{align}
 
\begin{equation}
B_{1}=\frac{3\kappa_{3}b^{3}E_{p}}{\left(\kappa_{2}+2\kappa_{3}\right)b^{3}+2\left(\kappa_{2}-\kappa_{3}\right)a^{3}},\label{eq:B1}
\end{equation}
 and 
\begin{equation}
B_{\ell}=0\;\textnormal{ for }\;\ell\geq2.\label{eq:B_ell}
\end{equation}
 With coefficients $B_{0},$ $B_{1},$ and $B_{\ell\geq2}$ defined,
Eqs. (\ref{eq:V2-10}) and (\ref{eq:V3-20}) become 
\begin{align}
V_{2}\left(r,\theta\right) & \approx B_{0}\left(1-\frac{a}{r}\right)+\frac{aV_{0}}{r}\nonumber \\
 & +B_{1}\left(1-\frac{a^{3}}{r^{3}}\right)r\cos\theta\label{eq:V2-20}
\end{align}
 and 
\begin{align}
V_{3}\left(r,\theta\right) & \approx E_{p}\left(h-s\right)+V_{L}+E_{p}r\cos\theta\nonumber \\
 & +\left[B_{0}\left(b-a\right)+aV_{0}-bE_{p}\left(h-s\right)-bV_{L}\right]\frac{1}{r}\nonumber \\
 & +\left[B_{1}\left(b^{3}-a^{3}\right)-b^{3}E_{p}\right]\frac{\cos\theta}{r^{2}},\label{eq:V3-30}
\end{align}
 where $V_{0}$ is the only unknown. 

The $V_{0}$ is evaluated from the statement about the discontinuity
of electric displacement at $r=a.$ At $r=a,$  the normal component
of the electric displacement suffers a discontinuity given by 
\begin{equation}
\left.\left[\mathbf{e}_{r}\cdot\mathbf{D}_{2}\left(r,\theta\right)-\mathbf{e}_{r}\cdot\mathbf{D}_{1}\left(r,\theta\right)\right]\right|_{r=a}=\sigma_{1},\label{eq:continuity-of-D-at-r=00003Da}
\end{equation}
 where $\sigma_{1}$ is the surface free-charge density at $r=a$
and $\mathbf{D}_{1}$ is the electric displacement in region $M_{1}.$
Repeating the same procedure outlined from Eq. (\ref{eq:continuity-of-D-at-r=00003Db})
through Eq. (\ref{eq:Neumann-BC-r=00003Db}), it can be shown that
\[
\left.\left[\kappa_{1}\frac{\partial V_{1}\left(r,\theta\right)}{\partial r}-\kappa_{2}\frac{\partial V_{2}\left(r,\theta\right)}{\partial r}\right]\right|_{r=a}=\frac{\sigma_{1}}{\epsilon_{0}}.
\]
 Since region $M_{1}$ is a conductor, 
\begin{align*}
\frac{\partial V_{1}\left(r,\theta\right)}{\partial r} & =0
\end{align*}
 and Neumann boundary condition at $r=a$ becomes 
\begin{equation}
\left.\frac{\partial V_{2}\left(r,\theta\right)}{\partial r}\right|_{r=a}=-\frac{\sigma_{1}}{\epsilon_{0}\kappa_{2}}.\label{eq:Neumann-BC-r=00003Da}
\end{equation}
 Using the results in Eqs. (\ref{eq:B0-pre}), (\ref{eq:B1}), and
(\ref{eq:B_ell}), the derivative in Eq. (\ref{eq:Neumann-BC-r=00003Da})
is readily computed from Eq. (\ref{eq:derivative-V2}), 
\[
\frac{\partial V_{2}\left(r,\theta\right)}{\partial r}=\left(B_{0}-V_{0}\right)\frac{a}{r^{2}}+B_{1}\left(1+\frac{2a^{3}}{r^{3}}\right)\cos\theta.
\]
 With this result, Eq. (\ref{eq:Neumann-BC-r=00003Da}) becomes 
\begin{equation}
\frac{B_{0}-V_{0}}{a}+3B_{1}\cos\theta=-\frac{\sigma_{1}}{\epsilon_{0}\kappa_{2}}.\label{eq:sigma1}
\end{equation}
 The $\cos\theta$ in Eq. (\ref{eq:sigma1}) can be eliminated by
integrating both sides over the spherical surface at $r=a,$ 
\begin{align*}
\int_{\theta=0}^{\pi}\int_{\phi=0}^{2\pi}\left(\frac{B_{0}-V_{0}}{a}+3B_{1}\cos\theta\right)a^{2}\sin\theta d\theta d\phi\\
=-\int_{\theta=0}^{\pi}\int_{\phi=0}^{2\pi}\frac{\sigma_{1}}{\epsilon_{0}\kappa_{2}}a^{2}\sin\theta d\theta d\phi,
\end{align*}
 yielding 
\begin{equation}
B_{0}-V_{0}=-\frac{a\sigma_{1}}{\epsilon_{0}\kappa_{2}}.\label{eq:B0-V0-explained}
\end{equation}
 What I have just done here only surmounts to the computing of total
free charge on the sphere of radius $r=a.$ For instance, in Eq. (\ref{eq:sigma1}),
one can integrate both sides over the surface $r=a$ of a sphere.
The right hand side yields total free charge on the surface $r=a,$
ignoring the extra constant factor. The left hand side yields terms
with $\cos\theta$ eliminated, as this term has been integrated over.
Canceling out the common terms yields Eq. (\ref{eq:B0-V0-explained}). 

That explained, $B_{0}$ is inserted from Eq. (\ref{eq:B0-pre}) into
Eq. (\ref{eq:B0-V0-explained}) to solve for $V_{0};$ and, this yields
\begin{align}
V_{0} & =V_{L}+\frac{a\left(b-a\right)\sigma_{1}}{b\epsilon_{0}\kappa_{2}}+\frac{a^{2}\sigma_{1}+b^{2}\sigma_{2}}{b\epsilon_{0}\kappa_{3}}\nonumber \\
 & +E_{p}\left(h-s\right).\label{eq:V0}
\end{align}
 With Eq. (\ref{eq:V0}), the coefficient $B_{0}$ of Eq. (\ref{eq:B0-pre})
becomes 

\begin{align}
B_{0} & =V_{L}+\frac{a\left(2b-a\right)\sigma_{1}}{b\epsilon_{0}\kappa_{2}}+\frac{a^{2}\sigma_{1}+b^{2}\sigma_{2}}{b\epsilon_{0}\kappa_{3}}\nonumber \\
 & +E_{p}\left(h-s\right).\label{eq:B0}
\end{align}
 With coefficients $B_{1},$ $V_{0},$ and $B_{0}$ defined respectively
in Eqs. (\ref{eq:B1}), (\ref{eq:V0}), and (\ref{eq:B0}), the electrostatic
potentials for regions $M_{1},$ $M_{2},$ and $M_{3}$ are obtained
from Eqs. (\ref{eq:V1-10}), (\ref{eq:V2-20}), and (\ref{eq:V3-30}).
They are 
\begin{align}
V_{1} & =V_{L}+\alpha+E_{p}\left(h-s\right),\quad r\leq a,\label{eq:V1-FINAL}
\end{align}

\begin{align}
V_{2}\left(r,\theta\right) & =V_{L}+\beta+E_{p}\left(h-s+\gamma r\cos\theta\right)\nonumber \\
 & -\frac{\lambda}{r}-\frac{a^{3}\gamma E_{p}\cos\theta}{r^{2}},\quad a<r\leq b,\label{eq:V2-FINAL}
\end{align}
 and  
\begin{align}
 & V_{3}\left(r,\theta\right)=V_{L}+E_{p}\left(h-s+r\cos\theta\right)+\frac{\nu}{r}\nonumber \\
 & \quad+\frac{\left[\gamma\left(b^{3}-a^{3}\right)-b^{3}\right]E_{p}\cos\theta}{r^{2}}+C,\quad r>b,\label{eq:V3-FINAL}
\end{align}
 where $\alpha,$ $\beta,$ $\gamma,$ $\lambda,$ and $\nu$ are
defined as 
\begin{align}
\alpha & =\frac{a\left(b-a\right)\sigma_{1}}{b\epsilon_{0}\kappa_{2}}+\frac{a^{2}\sigma_{1}+b^{2}\sigma_{2}}{b\epsilon_{0}\kappa_{3}},\nonumber \\
\beta & =\frac{a\left(2b-a\right)\sigma_{1}}{b\epsilon_{0}\kappa_{2}}+\frac{a^{2}\sigma_{1}+b^{2}\sigma_{2}}{b\epsilon_{0}\kappa_{3}},\nonumber \\
\gamma & =\frac{3\kappa_{3}b^{3}}{\left(\kappa_{2}+2\kappa_{3}\right)b^{3}+2\left(\kappa_{2}-\kappa_{3}\right)a^{3}},\label{eq:alpha-beta-gamma-lambda-nu}\\
\lambda & =\frac{a^{2}\sigma_{1}}{\epsilon_{0}\kappa_{2}},\nonumber \\
\nu & =\frac{2a\left(b-a\right)\sigma_{1}}{\epsilon_{0}\kappa_{2}}+\frac{a^{2}\sigma_{1}+b^{2}\sigma_{2}}{\epsilon_{0}\kappa_{3}}.\nonumber 
\end{align}
 For all of the treatment hereafter, only the derivatives of $V_{3},$
in particular, the normal derivatives associated with the plane-parallel
plate electrodes, are of importance. Therefore, the explicit expression
for the constant $C$ in Eq. (\ref{eq:V3-FINAL}) is not of much concern
here.

\subsection{Induced surface charges on conductor plates}

In spherical polar coordinate system, $\nabla$ operator  is defined
 by 
\[
\nabla=\mathbf{e}_{r}\frac{\partial}{\partial r}+\mathbf{e}_{\theta}\frac{1}{r}\frac{\partial}{\partial\theta}+\mathbf{e}_{\phi}\frac{1}{r\sin\theta}\frac{\partial}{\partial\phi},
\]
 where 
\begin{align*}
\mathbf{e}_{r} & =\mathbf{e}_{x}\sin\theta\cos\phi+\mathbf{e}_{y}\sin\theta\sin\phi+\mathbf{e}_{z}\cos\theta,\\
\mathbf{e}_{\theta} & =\mathbf{e}_{x}\cos\theta\cos\phi+\mathbf{e}_{y}\cos\theta\sin\phi-\mathbf{e}_{z}\sin\theta,\\
\mathbf{e}_{\phi} & =-\mathbf{e}_{x}\sin\phi+\mathbf{e}_{y}\cos\phi.
\end{align*}
 Hence, the $\mathbf{e}_{z}$ component of $\nabla$ operator is given
by 
\[
\mathbf{e}_{z}\left(\mathbf{e}_{z}\cdot\nabla\right)=\mathbf{e}_{z}\cos\theta\frac{\partial}{\partial r}-\mathbf{e}_{z}\frac{\sin\theta}{r}\frac{\partial}{\partial\theta}.
\]
 Using the form defined in Eq. (\ref{eq:electric-displacement-definition}),
the electric displacement in region $M_{3}$ is given by 
\[
\mathbf{D}_{3}\left(r,\theta\right)=-\epsilon_{0}\kappa_{3}\nabla V_{3}\left(r,\theta\right).
\]
 The $\mathbf{e}_{z}$ component of $\mathbf{D}_{3}\left(r,\theta\right)$
is obtained by replacing the $\nabla$ with the $\mathbf{e}_{z}\left(\mathbf{e}_{z}\cdot\nabla\right)$
operator and this gives 
\[
\mathbf{D}_{3;z}\left(r,\theta\right)=\epsilon_{0}\kappa_{3}\mathbf{e}_{z}\left[\frac{\sin\theta}{r}\frac{\partial V_{3}\left(r,\theta\right)}{\partial\theta}-\cos\theta\frac{\partial V_{3}\left(r,\theta\right)}{\partial r}\right],
\]
 where the notation $\mathbf{D}_{3;z}\left(r,\theta\right)$ denotes
the $\mathbf{e}_{z}$ component of $\mathbf{D}_{3}\left(r,\theta\right).$
With $V_{3}\left(r,\theta\right)$ of Eq. (\ref{eq:V3-FINAL}), the
$\mathbf{e}_{z}$ component of electric displacement in region $M_{3}$
is given by 

\begin{align}
 & \mathbf{D}_{3;z}\left(r,\theta\right)=\epsilon_{0}\kappa_{3}\mathbf{e}_{z}\left\{ \vphantom{\frac{\frac{1}{1^{1}}}{\frac{1^{1}}{1}}}\frac{\nu}{r^{2}}\cos\theta\right.\nonumber \\
 & \quad\left.+\frac{\left[\gamma\left(b^{3}-a^{3}\right)-b^{3}\right]E_{p}}{r^{3}}\left(3\cos^{2}\theta-1\right)-E_{p}\vphantom{\frac{\frac{1}{1^{1}}}{\frac{1^{1}}{1}}}\right\} .\label{eq:D3-z-z-comp-10}
\end{align}

The surface of the upper conductor plate is described by the Cartesian
$z=s$ plane. In the spherical polar coordinate system, the surface
of the upper conductor plate is described by 
\[
\cos\theta=\frac{s}{\sqrt{x^{2}+y^{2}+s^{2}}}.
\]
 Insertion of the expression for $\cos\theta$ into Eq. (\ref{eq:D3-z-z-comp-10})
yields 
\begin{align}
 & \mathbf{D}_{3;z}\left(x,y,s\right)=\mathbf{e}_{z}\epsilon_{0}\kappa_{3}\left\{ \frac{3\left[\gamma\left(b^{3}-a^{3}\right)-b^{3}\right]E_{p}s^{2}}{\left(x^{2}+y^{2}+s^{2}\right)^{5/2}}\right.\nonumber \\
 & \left.+\frac{\nu s-\left[\gamma\left(b^{3}-a^{3}\right)-b^{3}\right]E_{p}}{\left(x^{2}+y^{2}+s^{2}\right)^{3/2}}-E_{p}\right\} .\label{eq:z-comp-UP}
\end{align}
 At the surface of the upper conductor plate, the electric displacement
suffers a discontinuity given by 
\begin{equation}
\mathbf{e}_{z}\cdot\mathbf{D}_{ucp;z}\left(x,y,s\right)-\mathbf{e}_{z}\cdot\mathbf{D}_{3;z}\left(x,y,s\right)=\sigma_{iup},\label{eq:discon-D-upper}
\end{equation}
 where $\sigma_{iup}$ is the induced surface charge density on the
surface of the upper conductor plate and $\mathbf{D}_{ucp;z}$ is
the $\mathbf{e}_{z}$ component of the electric displacement inside
of the upper conductor plate. Since the electric displacement inside
of the upper conductor plate is zero, Eq. (\ref{eq:discon-D-upper})
reduces to 
\[
\mathbf{e}_{z}\cdot\mathbf{D}_{3;z}\left(x,y,s\right)=-\sigma_{iup}
\]
 and the surface charge density is given by 
\begin{align}
\sigma_{iup} & =-\epsilon_{0}\kappa_{3}\left\{ \frac{3\left[\gamma\left(b^{3}-a^{3}\right)-b^{3}\right]E_{p}s^{2}}{\left(x^{2}+y^{2}+s^{2}\right)^{5/2}}\right.\nonumber \\
 & \left.+\frac{\nu s-\left[\gamma\left(b^{3}-a^{3}\right)-b^{3}\right]E_{p}}{\left(x^{2}+y^{2}+s^{2}\right)^{3/2}}-E_{p}\right\} ,\label{eq:sigma_iup}
\end{align}
 where Eq. (\ref{eq:z-comp-UP}) has been inserted for $\mathbf{D}_{3;z}\left(x,y,s\right).$

The surface of the lower conductor plate is described by the Cartesian
$z=s-h$ plane. In the spherical polar coordinate system, the surface
of the lower conductor plate is given by 
\[
\cos\theta=\frac{s-h}{\sqrt{x^{2}+y^{2}+\left(s-h\right)^{2}}}
\]
 and Eq. (\ref{eq:D3-z-z-comp-10}) becomes 
\begin{align}
 & \mathbf{D}_{3;z}\left(x,y,s-h\right)\nonumber \\
 & \quad=\mathbf{e}_{z}\epsilon_{0}\kappa_{3}\left\{ \frac{3\left[\gamma\left(b^{3}-a^{3}\right)-b^{3}\right]E_{p}\left(s-h\right)^{2}}{\left[x^{2}+y^{2}+\left(s-h\right)^{2}\right]^{5/2}}\right.\nonumber \\
 & \quad\left.+\frac{\nu\left(s-h\right)-\left[\gamma\left(b^{3}-a^{3}\right)-b^{3}\right]E_{p}}{\left[x^{2}+y^{2}+\left(s-h\right)^{2}\right]^{3/2}}-E_{p}\right\} .\label{eq:z-comp-LP}
\end{align}
 At the surface of the lower conductor plate, the electric displacement
suffers a discontinuity given by 
\begin{equation}
\mathbf{e}_{z}\cdot\mathbf{D}_{3;z}\left(x,y,s-h\right)-\mathbf{e}_{z}\cdot\mathbf{D}_{lcp;z}\left(x,y,s-h\right)=\sigma_{ilp},\label{eq:discon-D-lower}
\end{equation}
 where $\sigma_{ilp}$ is the induced surface charge density on the
surface of the lower conductor plate and $\mathbf{D}_{lcp;z}$ is
the $\mathbf{e}_{z}$ component of the electric displacement inside
of the lower conductor plate. Since the electric displacement inside
of the lower conductor plate is zero, Eq. (\ref{eq:discon-D-lower})
reduces to 
\[
\mathbf{e}_{z}\cdot\mathbf{D}_{3;z}\left(x,y,s-h\right)=\sigma_{ilp}
\]
 and, with Eq. (\ref{eq:z-comp-LP}) inserted for $\mathbf{D}_{3;z}\left(x,y,s-h\right),$
the induced surface charge density is given by 
\begin{align}
\sigma_{ilp} & =\epsilon_{0}\kappa_{3}\left\{ \frac{3\left[\gamma\left(b^{3}-a^{3}\right)-b^{3}\right]E_{p}\left(h-s\right)^{2}}{\left[x^{2}+y^{2}+\left(h-s\right)^{2}\right]^{5/2}}\right.\nonumber \\
 & \left.-\frac{\nu\left(h-s\right)+\left[\gamma\left(b^{3}-a^{3}\right)-b^{3}\right]E_{p}}{\left[x^{2}+y^{2}+\left(h-s\right)^{2}\right]^{3/2}}-E_{p}\right\} ,\label{eq:sigma_ilp}
\end{align}
 where $\left(s-h\right)$ has been re-expressed as $-\left(h-s\right)$
purely for convenience. 

In the limit the parallel plates become infinite in extent, the total
of induced charges on the surfaces of each conductor plates must add
up to the total charge carried by the particle. To check on this,
Eqs. (\ref{eq:sigma_iup}) and (\ref{eq:sigma_ilp}) are integrated
over the surfaces of infinite parallel conductor plates with gap $h.$
For convenience, I shall perform the integral in the polar coordinate
system. In terms of the polar coordinates, Eqs. (\ref{eq:sigma_iup})
and (\ref{eq:sigma_ilp}) become 
\begin{align}
\sigma_{iup}\left(\rho,s\right) & =-\epsilon_{0}\kappa_{3}\left\{ \frac{3\left[\gamma\left(b^{3}-a^{3}\right)-b^{3}\right]E_{p}s^{2}}{\left(\rho^{2}+s^{2}\right)^{5/2}}\right.\nonumber \\
 & \left.+\frac{\nu s-\left[\gamma\left(b^{3}-a^{3}\right)-b^{3}\right]E_{p}}{\left(\rho^{2}+s^{2}\right)^{3/2}}-E_{p}\right\} \label{eq:sigma_iup-PC}
\end{align}
 and 
\begin{align}
\sigma_{ilp}\left(\rho,s\right) & =\epsilon_{0}\kappa_{3}\left\{ \frac{3\left[\gamma\left(b^{3}-a^{3}\right)-b^{3}\right]E_{p}\left(h-s\right)^{2}}{\left[\rho^{2}+\left(h-s\right)^{2}\right]^{5/2}}\right.\nonumber \\
 & \left.-\frac{\nu\left(h-s\right)+\left[\gamma\left(b^{3}-a^{3}\right)-b^{3}\right]E_{p}}{\left[\rho^{2}+\left(h-s\right)^{2}\right]^{3/2}}-E_{p}\right\} ,\label{eq:sigma_ilp-PC}
\end{align}
 where $\rho\equiv\sqrt{x^{2}+y^{2}}.$ Since the surface in polar
coordinate system is symmetric about its axis, the total induced charges
on both conductors can be performed as follow: 
\begin{align*}
Q_{iT} & =Q_{iup}+Q_{ilp}\\
 & =\int_{\phi=0}^{2\pi}\int_{\rho=0}^{\infty}\left[\sigma_{iup}\left(\rho,s\right)+\sigma_{ilp}\left(\rho,s\right)\right]\rho d\rho d\phi\\
 & =2\pi\int_{\rho=0}^{\infty}\left[\sigma_{iup}\left(\rho,s\right)+\sigma_{ilp}\left(\rho,s\right)\right]\rho d\rho,
\end{align*}
 where $Q_{iup}$ and $Q_{ilp}$ are respectively the total induced
charge on the surface of the upper and the lower conductor plates.
With Eqs. (\ref{eq:sigma_iup-PC}) and (\ref{eq:sigma_ilp-PC}), the
$Q_{iT}$ becomes 
\begin{align}
\frac{Q_{iT}}{\pi\epsilon_{0}\kappa_{3}} & =-2\int_{0}^{\infty}\left\{ \vphantom{\frac{\frac{1^{1}}{\frac{1^{1}}{1}}}{\frac{\frac{1^{1}}{1}}{1^{1}}}}\frac{3\left[\gamma\left(b^{3}-a^{3}\right)-b^{3}\right]E_{p}s^{2}}{\left(\rho^{2}+s^{2}\right)^{5/2}}\right.\nonumber \\
 & -\frac{3\left[\gamma\left(b^{3}-a^{3}\right)-b^{3}\right]E_{p}\left(h-s\right)^{2}}{\left[\rho^{2}+\left(h-s\right)^{2}\right]^{5/2}}\nonumber \\
 & +\frac{\nu s-\left[\gamma\left(b^{3}-a^{3}\right)-b^{3}\right]E_{p}}{\left(\rho^{2}+s^{2}\right)^{3/2}}\nonumber \\
 & \left.+\frac{\nu\left(h-s\right)+\left[\gamma\left(b^{3}-a^{3}\right)-b^{3}\right]E_{p}}{\left[\rho^{2}+\left(h-s\right)^{2}\right]^{3/2}}\vphantom{\frac{\frac{1^{1}}{\frac{1^{1}}{1}}}{\frac{\frac{1^{1}}{1}}{1^{1}}}}\right\} \rho d\rho.\label{eq:Q-10}
\end{align}
 Equation (\ref{eq:Q-10}) involves the following integral types:
\begin{align}
\int_{0}^{\infty}\frac{\rho d\rho}{\left(\rho^{2}+c^{2}\right)^{3/2}} & =\left.-\frac{1}{\sqrt{\rho^{2}+c^{2}}}\right|_{0}^{\infty}=\frac{1}{c},\label{eq:INT-for-QiT-1}\\
\int_{0}^{\infty}\frac{\rho d\rho}{\left(\rho^{2}+c^{2}\right)^{5/2}} & =\left.-\frac{1}{3\left(\rho^{2}+c^{2}\right)^{3/2}}\right|_{0}^{\infty}=\frac{1}{3c^{3}}.\label{eq:INT-for-QiT-2}
\end{align}
 With the integral formulas of Eqs. (\ref{eq:INT-for-QiT-1}) and
(\ref{eq:INT-for-QiT-2}), the $Q_{iT}$ of Eq. (\ref{eq:Q-10}) is
integrated to yield 

\begin{align*}
Q_{iT} & =-4\pi\epsilon_{0}\kappa_{3}\nu.
\end{align*}
 Insertion of the explicit expression for $\nu$ from Eq. (\ref{eq:alpha-beta-gamma-lambda-nu})
yields 
\begin{align}
Q_{iT} & =-\left(Q_{b}+Q_{1}+Q_{2}\right),\label{eq:QiT}
\end{align}
 where 
\begin{align*}
Q_{b} & =8\pi a\left(b-a\right)\sigma_{1}\frac{\kappa_{3}}{\kappa_{2}},\\
Q_{1} & =4\pi a^{2}\sigma_{1},\\
Q_{2} & =4\pi b^{2}\sigma_{2}.
\end{align*}
 The three quantities are identified as follow. The $Q_{1}$ and $Q_{2}$
are the ``free charges'' on the surfaces at $r=a$ and $r=b,$ respectively.
The $Q_{b}$ is the charge contribution arising from the presence
of a dielectric shell surrounding the metallic core. This contribution
vanishes in the absence of free charge on metallic core (i.e., $\sigma_{1}=0$)
or dielectric shell (i.e., $b-a=0$).

\subsection{Particle dynamics }

Two major electrostatic forces are acting on the core-shell structured
charged-particle in Fig. \ref{fig:particle-in-capacitor}. One such
force is the electrostatic force between the induced charges on the
surface of the upper electrode plate and the charged-particle. This
force is denoted by $\mathbf{F}_{1}.$ The other force is the electrostatic
force arising between the induced charges on the surface of the lower
electrode plate and the charged-particle and this force is denoted
as $\mathbf{F}_{2}.$ The net force exerted on the charged-particle
by induced charges on each surfaces of the conductor plates is therefore
given by 
\begin{align}
\mathbf{F} & =\mathbf{F}_{1}+\mathbf{F}_{2}\nonumber \\
 & =-\frac{1}{2}Q_{T}\left(\underset{S_{1}}{\int}d\mathbf{E}_{1}+\underset{S_{2}}{\int}d\mathbf{E}_{2}\right),\label{eq:F-00}
\end{align}
 where $Q_{T}$ is the effective charge carried by the charged-particle
and $d\mathbf{E}_{1}$ and $d\mathbf{E}_{2}$ are respective differential
electric fields corresponding to the upper and lower electrode plate
surfaces $S_{1}$ and $S_{2},$ respectively. For instance, $d\mathbf{E}_{1}$
is the differential electric field associated with the induced surface
charge at location $\mathbf{R}_{1}$ in Fig. \ref{fig:particle-in-capacitor}(a).
Similarly, $d\mathbf{E}_{2}$ is the differential electric field associated
with the induced surface charge at location $\mathbf{R}_{2}$ of Fig.
\ref{fig:particle-in-capacitor}(a).

The presence of extra factor of $1/2,$ the negative sign, and the
exact form of $Q_{T}$ in Eq. (\ref{eq:F-00}) can be explained as
follow. The extra factor of $1/2$ in Eq. (\ref{eq:F-00}) comes from
the fact that each parallel conductor plates sees only an hemisphere
of the charged-particle. The effective charge carried by the particle
is identical in magnitude to the $Q_{iT}$ of Eq. (\ref{eq:QiT}),
but with opposite charge polarity. Thus, 
\begin{align*}
Q_{T} & =-Q_{iT}
\end{align*}
 or 
\begin{align}
Q_{T} & =8\pi a\left(b-a\right)\sigma_{1}\frac{\kappa_{3}}{\kappa_{2}}+4\pi\left(a^{2}\sigma_{1}+b^{2}\sigma_{2}\right).\label{eq:QT}
\end{align}
 The negative sign in Eq. (\ref{eq:F-00}) is necessary for specifying
correctly the direction of the forces exerted on the core-shell structured
charge-particle by induced surface charges from each parallel conductor
plates. To demonstrate this, the integrals in Eq. (\ref{eq:F-00})
can be represented by 

\begin{equation}
\underset{S_{i}}{\int}d\mathbf{E}_{i}\;\rightarrow\;\frac{1}{4\pi\epsilon_{3}}\int_{\phi_{i}=0}^{2\pi}\int_{\rho_{i}=0}^{\rho}\frac{\varsigma_{i}\mathbf{R}_{i}\rho_{i}d\rho_{i}d\phi_{i}}{\left(\mathbf{R}_{i}\cdot\mathbf{R}_{i}\right)^{3/2}},\label{eq:integral_dEi_form}
\end{equation}
 where $\varsigma_{i}$ is the induced surface charge at location
$\mathbf{R}_{i}$ (i.e., $i=1,2$) in Fig. \ref{fig:particle-in-capacitor}(a).
Now, suppose if $\varsigma_{1}$ is positive, then the direction of
$d\mathbf{E}_{1}$ must be in $-\mathbf{R}_{1},$ as it can be inspected
from Fig. \ref{fig:particle-in-capacitor}(a). On the other hand,
if $\varsigma_{1}$ is negative, then the direction of $d\mathbf{E}_{1}$
must be in $\mathbf{R}_{1}.$ The same argument can be said for those
involving $\varsigma_{2},$ $d\mathbf{E}_{2},$ and $\mathbf{R}_{2}.$
And, this explains the presence of negative sign in Eq. (\ref{eq:F-00}). 

That said, using the form defined in Eq. (\ref{eq:integral_dEi_form}),
the force expression of Eq. (\ref{eq:F-00}) becomes 
\begin{align}
\mathbf{F}_{i} & =-\frac{Q_{T}}{8\pi\epsilon_{3}}\int_{\phi_{i}=0}^{2\pi}\int_{\rho_{i}=0}^{\rho}\frac{\varsigma_{i}\mathbf{R}_{i}\rho_{i}d\rho_{i}d\phi_{i}}{\left(\mathbf{R}_{i}\cdot\mathbf{R}_{i}\right)^{3/2}},\label{eq:F-10}
\end{align}
 where $i=\left(1,2\right),$ $\varsigma_{1}\equiv\sigma_{iup}$ of
Eq. (\ref{eq:sigma_iup-PC}), $\varsigma_{2}\equiv\sigma_{ilp}$ of
Eq. (\ref{eq:sigma_ilp-PC}), and $\epsilon_{3}$ is the electric
permittivity of the region $M_{3}.$ The explicit expression for $\mathbf{R}_{i},$
which defines the position of the $\varsigma_{i}$ associated with
$dS_{i}$ as illustrated in Fig. \ref{fig:particle-in-capacitor}(a)
for $i=\left(1,2\right),$ are given by 
\begin{align}
\mathbf{R}_{1} & =\mathbf{e}_{x}\rho_{1}\cos\phi_{1}+\mathbf{e}_{y}\rho_{1}\sin\phi_{1}+\mathbf{e}_{z}s,\label{eq:R1}\\
\mathbf{R}_{2} & =\mathbf{e}_{x}\rho_{2}\cos\phi_{2}+\mathbf{e}_{y}\rho_{2}\sin\phi_{2}+\mathbf{e}_{z}\left(s-h\right),\label{eq:R2}
\end{align}
 where $h>s.$ 

The force exerted on the particle by the induced charge on the surface
of the upper conductor plate is obtained by inserting $\mathbf{R}_{1}$
of Eq. (\ref{eq:R1}) into Eq. (\ref{eq:F-10}). This yields 
\begin{align}
 & \mathbf{F}_{1}=-\frac{Q_{T}}{8\pi\epsilon_{3}}\int_{\phi_{1}=0}^{2\pi}\int_{\rho_{1}=0}^{\rho}\left[\mathbf{e}_{x}\frac{\varsigma_{1}\rho_{1}\cos\phi_{1}}{\left(\rho_{1}^{2}+s^{2}\right)^{3/2}}\right.\nonumber \\
 & \left.+\mathbf{e}_{y}\frac{\varsigma_{1}\rho_{1}\sin\phi_{1}}{\left(\rho_{1}^{2}+s^{2}\right)^{3/2}}+\mathbf{e}_{z}\frac{\varsigma_{1}s}{\left(\rho_{1}^{2}+s^{2}\right)^{3/2}}\right]\rho_{1}d\rho_{1}d\phi_{1}.\label{eq:F-20}
\end{align}
 The two terms in the integrand with $\cos\phi_{1}$ and $\sin\phi_{1}$
vanish when integrated over $d\phi.$ Thus, Eq. (\ref{eq:F-20}) reduces
to 
\begin{equation}
\mathbf{F}_{1}=-\mathbf{e}_{z}\frac{Q_{T}s}{4\epsilon_{3}}\int_{0}^{\rho}\frac{\sigma_{iup}\rho_{1}d\rho_{1}}{\left(\rho_{1}^{2}+s^{2}\right)^{3/2}},\label{eq:F1-00}
\end{equation}
 where $\varsigma_{1}\equiv\sigma_{iup}.$ Insertion of the explicit
expression for $\sigma_{iup}$ from Eq. (\ref{eq:sigma_iup-PC}) into
Eq. (\ref{eq:F1-00}) yields 
\begin{align}
\mathbf{F}_{1} & =\mathbf{e}_{z}\frac{Q_{T}s}{4}\int_{0}^{\rho}\left\{ \frac{3\left[\gamma\left(b^{3}-a^{3}\right)-b^{3}\right]E_{p}s^{2}}{\left(\rho_{1}^{2}+s^{2}\right)^{4}}\right.\nonumber \\
 & +\frac{\nu s-\left[\gamma\left(b^{3}-a^{3}\right)-b^{3}\right]E_{p}}{\left(\rho_{1}^{2}+s^{2}\right)^{3}}\nonumber \\
 & \left.-\frac{E_{p}}{\left(\rho_{1}^{2}+s^{2}\right)^{3/2}}\right\} \rho_{1}d\rho_{1}.\label{eq:F1-10}
\end{align}
 Equation (\ref{eq:F1-10}) involves the following type of integrals:
\begin{align}
\int_{0}^{\rho}\frac{\rho_{1}d\rho_{1}}{\left(\rho_{1}^{2}+s^{2}\right)^{4}} & =\frac{1}{6s^{6}}-\frac{1}{6\left(\rho^{2}+s^{2}\right)^{3}},\label{eq:INT-1}\\
\int_{0}^{\rho}\frac{\rho_{1}d\rho_{1}}{\left(\rho_{1}^{2}+s^{2}\right)^{3}} & =\frac{1}{4s^{4}}-\frac{1}{4\left(\rho^{2}+s^{2}\right)^{2}},\label{eq:INT-2}\\
\int_{0}^{\rho}\frac{\rho_{1}d\rho_{1}}{\left(\rho_{1}^{2}+s^{2}\right)^{3/2}} & =\frac{1}{s}-\frac{1}{\sqrt{\rho^{2}+s^{2}}}.\label{eq:INT-3}
\end{align}
 Insertion of Eqs. (\ref{eq:INT-1}), (\ref{eq:INT-2}), and (\ref{eq:INT-3})
into Eq. (\ref{eq:F1-10}) yields 
\begin{align}
\mathbf{F}_{1} & =\mathbf{e}_{z}\frac{Q_{T}}{16}\left\{ \frac{\nu}{s^{2}}-\frac{\nu s^{2}}{\left(\rho^{2}+s^{2}\right)^{2}}+\frac{\left[\gamma\left(b^{3}-a^{3}\right)-b^{3}\right]E_{p}}{s^{3}}\right.\nonumber \\
 & +\frac{\left[\gamma\left(b^{3}-a^{3}\right)-b^{3}\right]E_{p}s}{\left(\rho^{2}+s^{2}\right)^{2}}-\frac{2\left[\gamma\left(b^{3}-a^{3}\right)-b^{3}\right]E_{p}s^{3}}{\left(\rho^{2}+s^{2}\right)^{3}}\nonumber \\
 & \left.+\frac{4E_{p}s}{\sqrt{\rho^{2}+s^{2}}}-4E_{p}\right\} ,\label{eq:F1-20}
\end{align}
 where $b\leq s\leq h-b.$ Equation (\ref{eq:F1-20}) is the force
exerted on the charged-particle by the induced charges on the surface
of the upper conductor plate. 

The expression for the force exerted on the particle by the induced
charges on the surface of the lower conductor plate is obtained by
inserting $\mathbf{R}_{2}$ of Eq. (\ref{eq:R2}) into Eq. (\ref{eq:F-10}).
Repeating the similar procedure outlined in Eqs. (\ref{eq:F-20})
and (\ref{eq:F1-00}), one obtains 
\begin{equation}
\mathbf{F}_{2}=\mathbf{e}_{z}\frac{Q_{T}\left(h-s\right)}{4\epsilon_{3}}\int_{0}^{\rho}\frac{\sigma_{ilp}\rho_{2}d\rho_{2}}{\left[\rho_{2}^{2}+\left(h-s\right)^{2}\right]^{3/2}}.\label{eq:F2-00}
\end{equation}
 Insertion of the explicit expression for $\sigma_{ilp}$ from Eq.
(\ref{eq:sigma_ilp-PC}) into Eq. (\ref{eq:F2-00}) yields 
\begin{align}
\mathbf{F}_{2} & =\mathbf{e}_{z}\frac{Q_{T}\left(h-s\right)}{4}\int_{0}^{\rho}\left\{ \frac{3\left[\gamma\left(b^{3}-a^{3}\right)-b^{3}\right]E_{p}\left(h-s\right)^{2}}{\left[\rho_{2}^{2}+\left(h-s\right)^{2}\right]^{4}}\right.\nonumber \\
 & -\frac{\nu\left(h-s\right)+\left[\gamma\left(b^{3}-a^{3}\right)-b^{3}\right]E_{p}}{\left[\rho_{2}^{2}+\left(h-s\right)^{2}\right]^{3}}\nonumber \\
 & \left.-\frac{E_{p}}{\left[\rho_{2}^{2}+\left(h-s\right)^{2}\right]^{3/2}}\right\} \rho_{2}d\rho_{2}.\label{eq:F2-10}
\end{align}
 Using the integral formulas from Eqs. (\ref{eq:INT-1}), (\ref{eq:INT-2}),
and (\ref{eq:INT-3}) with $s$ replaced by $h-s,$ Eq. (\ref{eq:F2-10})
becomes 
\begin{align}
 & \mathbf{F}_{2}=\mathbf{e}_{z}\frac{Q_{T}}{16}\left\{ \frac{\nu\left(h-s\right)^{2}}{\left[\rho^{2}+\left(h-s\right)^{2}\right]^{2}}-\frac{\nu}{\left(h-s\right)^{2}}\right.\nonumber \\
 & +\frac{\left[\gamma\left(b^{3}-a^{3}\right)-b^{3}\right]E_{p}}{\left(h-s\right)^{3}}+\frac{\left[\gamma\left(b^{3}-a^{3}\right)-b^{3}\right]E_{p}\left(h-s\right)}{\left[\rho^{2}+\left(h-s\right)^{2}\right]^{2}}\nonumber \\
 & -\frac{\left[\gamma\left(b^{3}-a^{3}\right)-b^{3}\right]E_{p}\left(h-s\right)^{3}}{\left[\rho^{2}+\left(h-s\right)^{2}\right]^{3}}\nonumber \\
 & \left.+\frac{4E_{p}\left(h-s\right)}{\sqrt{\rho^{2}+\left(h-s\right)^{2}}}-4E_{p}\vphantom{\frac{\nu\left(h-s\right)^{2}}{\left[\rho^{2}+\left(h-s\right)^{2}\right]^{2}}}\right\} ,\label{eq:F2-20}
\end{align}
 where $b\leq s\leq h-b.$ Equation (\ref{eq:F2-20}) is the force
exerted on the charged-particle by the induced charges on the surface
of the lower conductor plate. 

For a  parallel plate system, which is microscopically large, but
macroscopically small, the forces in Eqs. (\ref{eq:F1-20}) and (\ref{eq:F2-20})
can be approximated by making $\rho$ go to infinity. This approximation
is certainly valid for very small charged-particles confined between
large parallel conductor plates. In the limit $\rho$ goes to infinity,
Eqs. (\ref{eq:F1-20}) and (\ref{eq:F2-20}) simplify in form as 

\begin{align}
\mathbf{F}_{1} & =\mathbf{e}_{z}\frac{Q_{T}}{4}\left\{ \frac{\nu}{4s^{2}}+\frac{\left[\gamma\left(b^{3}-a^{3}\right)-b^{3}\right]E_{p}}{4s^{3}}-E_{p}\right\} \label{eq:F1-20-infinite}
\end{align}
 and 
\begin{align}
\mathbf{F}_{2} & =\mathbf{e}_{z}\frac{Q_{T}}{4}\left\{ \frac{\left[\gamma\left(b^{3}-a^{3}\right)-b^{3}\right]E_{p}}{4\left(h-s\right)^{3}}-\frac{\nu}{4\left(h-s\right)^{2}}-E_{p}\right\} ,\label{eq:F2-20-infinite}
\end{align}
 where $b\leq s\leq h-b$ and $\nu$ is defined in Eq. (\ref{eq:alpha-beta-gamma-lambda-nu}),
\begin{align*}
\nu & =\frac{2a\left(b-a\right)\sigma_{1}}{\epsilon_{0}\kappa_{2}}+\frac{a^{2}\sigma_{1}+b^{2}\sigma_{2}}{\epsilon_{0}\kappa_{3}}.
\end{align*}
 Notice that the resulting forces in Eqs. (\ref{eq:F1-20-infinite})
and (\ref{eq:F2-20-infinite}) are now just one dimensional forces;
that is, $\mathbf{F}_{1}\equiv\mathbf{F}_{1}\left(s\right)$ and $\mathbf{F}_{2}\equiv\mathbf{F}_{2}\left(s\right),$
where the parameter $s$ is the relative distance between the center
of mass point of core-shell structured charged-particle and the surface
of upper conductor plate. The dynamics of charged-particle system
illustrated in Fig. \ref{fig:particle-in-capacitor} has now reduced
down to solving a nonlinear ordinary differential equation. 

Insertion of Eqs. (\ref{eq:F1-20-infinite}) and (\ref{eq:F2-20-infinite})
into Eq. (\ref{eq:F-00}) yields the total force exerted on the charged-particle
by the induced charges on the surfaces of parallel plate conductors.
The result is 
\begin{align*}
\mathbf{F} & =\mathbf{e}_{z}\frac{Q_{T}}{16}\left\{ \frac{\nu}{s^{2}}-\frac{\nu}{\left(h-s\right)^{2}}+\frac{\left[\gamma\left(b^{3}-a^{3}\right)-b^{3}\right]E_{p}}{s^{3}}\right.\\
 & \left.+\frac{\left[\gamma\left(b^{3}-a^{3}\right)-b^{3}\right]E_{p}}{\left(h-s\right)^{3}}-8E_{p}\right\} .
\end{align*}
 If the gravitational effect is included, the force experienced by
the particle is 
\begin{align*}
\mathbf{F}_{T} & =\mathbf{F}-\mathbf{e}_{z}mg
\end{align*}
 or 
\begin{align}
\mathbf{F}_{T} & =\mathbf{e}_{z}\frac{Q_{T}}{16}\left\{ \frac{\nu}{s^{2}}-\frac{\nu}{\left(h-s\right)^{2}}+\frac{\left[\gamma\left(b^{3}-a^{3}\right)-b^{3}\right]E_{p}}{s^{3}}\right.\nonumber \\
 & \left.+\frac{\left[\gamma\left(b^{3}-a^{3}\right)-b^{3}\right]E_{p}}{\left(h-s\right)^{3}}-8E_{p}\right\} -\mathbf{e}_{z}mg,\label{eq:F-FINAL-pre}
\end{align}
 where $m$ is the mass of the particle, $g=9.8\,\textnormal{m}\cdot\textnormal{s}^{-2}$
is the gravity constant, and the gravitational force has been assumed
to be in the $-\mathbf{e}_{z}$ direction. Since $Q_{T}$ is related
to $\nu$ by 
\begin{equation}
Q_{T}=4\pi\epsilon_{0}\kappa_{3}\nu,\label{eq:QT-in-terms-of-nu}
\end{equation}
 the $\mathbf{F}_{T}$ of Eq. (\ref{eq:F-FINAL-pre}) may be re-expressed
for convenience as 
\begin{align}
\mathbf{F}_{T} & =\mathbf{e}_{z}\frac{\pi\epsilon_{0}\kappa_{3}\nu}{4}\left\{ \frac{\nu}{s^{2}}-\frac{\nu}{\left(h-s\right)^{2}}+\frac{\left[\gamma\left(b^{3}-a^{3}\right)-b^{3}\right]E_{p}}{s^{3}}\right.\nonumber \\
 & \left.+\frac{\left[\gamma\left(b^{3}-a^{3}\right)-b^{3}\right]E_{p}}{\left(h-s\right)^{3}}-8E_{p}\right\} -\mathbf{e}_{z}mg.\label{eq:F-FINAL}
\end{align}

The dynamics of oscillating charged-particle is given by 

\begin{equation}
\mathbf{e}_{z}\frac{d}{dt}\left(\frac{mv}{\sqrt{1-\frac{v^{2}}{c^{2}}}}\right)=\mathbf{F}_{T},\label{eq:relativistic-LAW}
\end{equation}
 where $c=3\times10^{8}\,\textnormal{m}\cdot\textnormal{s}^{-1}$
is the speed of light in vacuum. The left hand side of Eq. (\ref{eq:relativistic-LAW})
can be differentiated to give 
\begin{equation}
\mathbf{e}_{z}v\frac{d}{dt}\left(\frac{1}{\sqrt{1-\frac{v^{2}}{c^{2}}}}\right)+\frac{\dot{v}\mathbf{e}_{z}}{\sqrt{1-\frac{v^{2}}{c^{2}}}}=\frac{\mathbf{F}_{T}}{m},\label{eq:rela-LAW-10}
\end{equation}
 where $\dot{v}\equiv dv/dt.$ Knowing that 
\[
\frac{d}{dt}\left(\frac{1}{\sqrt{1-\frac{v^{2}}{c^{2}}}}\right)=\frac{v\dot{v}}{c^{2}\left(1-\frac{v^{2}}{c^{2}}\right)^{3/2}},
\]
 equation  (\ref{eq:rela-LAW-10}) becomes 
\[
\frac{v^{2}\dot{v}\mathbf{e}_{z}}{c^{2}\left(1-\frac{v^{2}}{c^{2}}\right)^{3/2}}+\frac{\dot{v}\mathbf{e}_{z}}{\sqrt{1-\frac{v^{2}}{c^{2}}}}=\frac{\mathbf{F}_{T}}{m}.
\]
 Multiplying both sides by $c^{2}\left(1-v^{2}/c^{2}\right)^{3/2}$
yields 
\begin{equation}
\dot{v}\mathbf{e}_{z}=\frac{\mathbf{F}_{T}}{m}\left(1-\frac{v^{2}}{c^{2}}\right)^{3/2}.\label{eq:dvdt-relativity}
\end{equation}
Since $v=\dot{s}$ and $\dot{v}=\ddot{s},$ Eq. (\ref{eq:dvdt-relativity})
becomes 
\[
\ddot{s}\mathbf{e}_{z}=\frac{\mathbf{F}_{T}}{m}\left(1-\frac{\dot{s}^{2}}{c^{2}}\right)^{3/2},
\]
 where $\ddot{s}\equiv d^{2}s/dt^{2}.$ With $\mathbf{F}_{T}$ explicitly
inserted from Eq. (\ref{eq:F-FINAL}), the expression for $\ddot{s}$
becomes 
\begin{align}
\ddot{s} & =\left(1-\frac{\dot{s}^{2}}{c^{2}}\right)^{3/2}\left(\frac{\pi\epsilon_{0}\kappa_{3}\nu}{4m}\left\{ \frac{\nu}{s^{2}}-\frac{\nu}{\left(h-s\right)^{2}}\right.\right.\nonumber \\
 & +\frac{\left[\gamma\left(b^{3}-a^{3}\right)-b^{3}\right]E_{p}}{s^{3}}\nonumber \\
 & \left.\left.+\frac{\left[\gamma\left(b^{3}-a^{3}\right)-b^{3}\right]E_{p}}{\left(h-s\right)^{3}}-8E_{p}\right\} -g\right),\label{eq:ODE-rela-in-s}
\end{align}
 where $\mathbf{e}_{z}$ has been dropped for convenience. It is convenient
to re-express Eq. (\ref{eq:ODE-rela-in-s}) in terms of the variable
$z_{d}$ illustrated in Fig. \ref{fig:particle-in-capacitor}(a).
Two variables, $s$ and $z_{d},$ are related to each other by 
\begin{equation}
s=z_{d}+b,\quad\dot{s}=\dot{z}_{d},\quad\ddot{s}=\ddot{z}_{d},\label{eq:s-in-zd}
\end{equation}
 where $b$ is a constant. Hence, in terms of $z_{d},$ Eq. (\ref{eq:ODE-rela-in-s})
becomes 
\begin{align}
\ddot{z}_{d} & =\left(1-\frac{\dot{z}_{d}^{2}}{c^{2}}\right)^{3/2}\left(\frac{\pi\epsilon_{0}\kappa_{3}\nu}{4m}\left\{ \frac{\nu}{\left(z_{d}+b\right)^{2}}\right.\right.\nonumber \\
 & -\frac{\nu}{\left(h-z_{d}-b\right)^{2}}+\frac{\left[\gamma\left(b^{3}-a^{3}\right)-b^{3}\right]E_{p}}{\left(z_{d}+b\right)^{3}}\nonumber \\
 & \left.\left.+\frac{\left[\gamma\left(b^{3}-a^{3}\right)-b^{3}\right]E_{p}}{\left(h-z_{d}-b\right)^{3}}-8E_{p}\right\} -g\right).\label{eq:ODE-rela}
\end{align}
 Equation (\ref{eq:ODE-rela}) governs the dynamics of an oscillating
charged-particle, subjected to high electrostatic fields, at all speeds.

\subsection{Criterion for charged-particle oscillation in the absence of charge
transfer process }

The criterion for charged-particle oscillation in the absence of charge
transfer process between the rebounding electrode and the charged-particle
can be obtained by analyzing the force expression of Eq. (\ref{eq:F-FINAL}), 

\begin{align}
\mathbf{F}_{T} & =\mathbf{e}_{z}\frac{\pi\epsilon_{0}\kappa_{3}\nu}{4}\left\{ \frac{\nu}{s^{2}}-\frac{\nu}{\left(h-s\right)^{2}}\right.\nonumber \\
 & +\frac{\left[\gamma\left(b^{3}-a^{3}\right)-b^{3}\right]E_{p}}{s^{3}}\nonumber \\
 & \left.+\frac{\left[\gamma\left(b^{3}-a^{3}\right)-b^{3}\right]E_{p}}{\left(h-s\right)^{3}}-8E_{p}\right\} -\mathbf{e}_{z}mg,\label{eq:F-OSC-1}
\end{align}
 The kinematics of charged-particle motion associated with the force
$\mathbf{F}_{T}$ of Eq. (\ref{eq:F-OSC-1}) is illustrated in Fig.
\ref{fig:zd(t)-plot}, where the parameters $s$ and $z_{d}$ are
related by $s=z_{d}+b.$ For the plot illustrated in Fig. \ref{fig:zd(t)-plot},
the turning points of charged-particle motion occurs approximately
at $z_{d}=z_{d,m}=0.25h.$ 

By definition, when the particle is in vicinity of the turning point,
but not past it, the net force acting on the particle is directed
in the opposite direction of particle's motion. Immediately past the
turning point, the particle's motion is in the same direction as the
net force. Thus, immediately past the turning point, the net force
satisfies 
\[
\mathbf{F}_{T}\left(z_{d,m}\right)>0
\]
 or 
\begin{align*}
\frac{\pi\epsilon_{0}\kappa_{3}\nu}{4}\left\{ \frac{\nu}{z_{d,m}^{2}}-\frac{\nu}{\left(h-z_{d,m}\right)^{2}}+\frac{\left[\gamma\left(b^{3}-a^{3}\right)-b^{3}\right]E_{p}}{z_{d,m}^{3}}\right.\\
\left.+\frac{\left[\gamma\left(b^{3}-a^{3}\right)-b^{3}\right]E_{p}}{\left(h-z_{d,m}\right)^{3}}-8E_{p}\right\} -mg>0.
\end{align*}
 Referring to the plot in Fig. \ref{fig:zd(t)-plot}, this result
simply states that net force is in $\mathbf{e}_{z}$ direction at
$z_{d}=z_{d,m}.$ This relation can be rewritten as 
\begin{align*}
\overbrace{\left[\frac{\gamma\left(b^{3}-a^{3}\right)-b^{3}}{z_{d,m}^{3}}+\frac{\gamma\left(b^{3}-a^{3}\right)-b^{3}}{\left(h-z_{d,m}\right)^{3}}-8\right]}^{\psi}E_{p}\\
>\frac{\nu}{\left(h-z_{d,m}\right)^{2}}-\frac{\nu}{z_{d,m}^{2}}+\frac{4mg}{\pi\epsilon_{0}\kappa_{3}\nu},
\end{align*}
 where 
\[
\gamma=\frac{3\kappa_{3}b^{3}}{\left(\kappa_{2}+2\kappa_{3}\right)b^{3}+2\left(\kappa_{2}-\kappa_{3}\right)a^{3}}
\]
 and 
\begin{equation}
\psi=\frac{\gamma\left(b^{3}-a^{3}\right)-b^{3}}{z_{d,m}^{3}}+\frac{\gamma\left(b^{3}-a^{3}\right)-b^{3}}{\left(h-z_{d,m}\right)^{3}}-8.\label{eq:psi}
\end{equation}
 Notice that $0<\gamma<1;$ therefore, $\psi<0.$ Thus, the previous
relation can be rewritten as 
\[
-\frac{\nu}{\left(h-z_{d,m}\right)^{2}}+\frac{\nu}{z_{d,m}^{2}}-\frac{4mg}{\pi\epsilon_{0}\kappa_{3}\nu}>-\psi E_{p}=\left|\psi\right|E_{p}.
\]
 Solving for $E_{p},$ I obtain 
\[
E_{p}<\frac{1}{\left|\psi\right|}\left[\frac{\nu}{z_{d,m}^{2}}-\frac{\nu}{\left(h-z_{d,m}\right)^{2}}-\frac{4mg}{\pi\epsilon_{0}\kappa_{3}\nu}\right].
\]
 In terms of the effective charge $Q_{T}$ defined in Eq. (\ref{eq:QT-in-terms-of-nu}),
this result becomes 
\begin{align}
E_{p} & <\frac{1}{Q_{T}\left|\psi\right|}\left\{ \frac{Q_{T}^{2}}{4\pi\epsilon_{0}\kappa_{3}}\left[\frac{1}{z_{d,m}^{2}}-\frac{1}{\left(h-z_{d,m}\right)^{2}}\right]-16mg\right\} .\label{eq:Ep-OSC-FINAL}
\end{align}
 By definition, $E_{p}\geq0$ because it is the magnitude of electric
field. Therefore, the right hand side of Eq. (\ref{eq:Ep-OSC-FINAL})
must be positive. But, according to the plot illustrated in Fig. \ref{fig:zd(t)-plot},
$z_{d,m}<h/2;$ and, thus 
\[
\frac{1}{z_{d,m}^{2}}>\frac{1}{\left(h-z_{d,m}\right)^{2}}.
\]
 For a positive particle, $Q_{T}>0$ and Eq. (\ref{eq:Ep-OSC-FINAL})
can always be satisfied for a sufficiently ionized positive core-shell
structured particle. For convenience, I shall express $Q_{T}$ in
terms of $E_{p}.$ Equation (\ref{eq:Ep-OSC-FINAL}) can be rearranged
to yield 
\[
\frac{Q_{T}^{2}}{4\pi\epsilon_{0}\kappa_{3}}\left[\frac{1}{z_{d,m}^{2}}-\frac{1}{\left(h-z_{d,m}\right)^{2}}\right]-\left|Q_{T}\right|\left|\psi\right|E_{p}-16mg>0.
\]
 Utilizing the quadratic formula, this can be solved for $Q_{T}$
to yield 
\begin{equation}
Q_{T}>\frac{2\pi\epsilon_{0}\kappa_{3}}{\xi}\left(\left|\psi\right|E_{p}+\sqrt{\psi^{2}E_{p}^{2}+\frac{16mg\xi}{\pi\epsilon_{0}\kappa_{3}}}\right),\label{eq:QT-OSC-FINAL}
\end{equation}
 where 
\begin{equation}
\xi=\frac{1}{z_{d,m}^{2}}-\frac{1}{\left(h-z_{d,m}\right)^{2}}\label{eq:Ehi}
\end{equation}
 and 
\[
\psi=\frac{\gamma\left(b^{3}-a^{3}\right)-b^{3}}{z_{d,m}^{3}}+\frac{\gamma\left(b^{3}-a^{3}\right)-b^{3}}{\left(h-z_{d,m}\right)^{3}}-8.
\]

When the core-shell structured charged-particle is negatively charged,
how does the charged-particle oscillation criterion get modified?
For the negatively charged core-shell structured particle, $\nu<0,$
where 
\[
\nu=\frac{2a\left(b-a\right)\sigma_{1}}{\epsilon_{0}\kappa_{2}}+\frac{a^{2}\sigma_{1}+b^{2}\sigma_{2}}{\epsilon_{0}\kappa_{3}}.
\]
Thus, the $\mathbf{F}_{T}$ of Eq. (\ref{eq:F-OSC-1}) becomes 
\begin{align*}
\mathbf{F}_{T} & =\mathbf{e}_{z}\frac{\pi\epsilon_{0}\kappa_{3}\left|\nu\right|}{4}\left\{ \frac{\left|\nu\right|}{s^{2}}-\frac{\left|\nu\right|}{\left(h-s\right)^{2}}\right.\\
 & -\frac{\left[\gamma\left(b^{3}-a^{3}\right)-b^{3}\right]E_{p}}{s^{3}}\\
 & \left.-\frac{\left[\gamma\left(b^{3}-a^{3}\right)-b^{3}\right]E_{p}}{\left(h-s\right)^{3}}+8E_{p}\right\} -\mathbf{e}_{z}mg.
\end{align*}
 Furthermore, since $\gamma\left(b^{3}-a^{3}\right)-b^{3}<0,$ I have
\begin{align}
\mathbf{F}_{T} & =\mathbf{e}_{z}\frac{\pi\epsilon_{0}\kappa_{3}\left|\nu\right|}{4}\left\{ \frac{\left|\nu\right|}{s^{2}}-\frac{\left|\nu\right|}{\left(h-s\right)^{2}}\right.\nonumber \\
 & +\frac{\left|\gamma\left(b^{3}-a^{3}\right)-b^{3}\right|E_{p}}{s^{3}}\nonumber \\
 & \left.+\frac{\left|\gamma\left(b^{3}-a^{3}\right)-b^{3}\right|E_{p}}{\left(h-s\right)^{3}}+8E_{p}\right\} -\mathbf{e}_{z}mg.\label{eq:F-OSC-2}
\end{align}
 Now, immediately past the turning point at $z_{d,m}=7.5\times10^{-4}\,\textnormal{m}$
in Fig. \ref{fig:zd(t)-plot-nC}, the net force acting on the particle
satisfies the condition given by 
\[
\mathbf{F}_{T}\left(z_{d,m}\right)<0
\]
 or 
\begin{align*}
 & \frac{\pi\epsilon_{0}\kappa_{3}\left|\nu\right|}{4}\left\{ \frac{\left|\nu\right|}{z_{d,m}^{2}}-\frac{\left|\nu\right|}{\left(h-z_{d,m}\right)^{2}}+\frac{\left|\gamma\left(b^{3}-a^{3}\right)-b^{3}\right|E_{p}}{z_{d,m}^{3}}\right.\\
 & \left.+\frac{\left|\gamma\left(b^{3}-a^{3}\right)-b^{3}\right|E_{p}}{\left(h-z_{d,m}\right)^{3}}+8E_{p}\right\} -mg<0,
\end{align*}
 where Eq. (\ref{eq:F-OSC-2}) has been substituted in for the explicit
expression. This relation is rearranged as 
\begin{align*}
 & \left[\frac{\left|\gamma\left(b^{3}-a^{3}\right)-b^{3}\right|}{z_{d,m}^{3}}+\frac{\left|\gamma\left(b^{3}-a^{3}\right)-b^{3}\right|}{\left(h-z_{d,m}\right)^{3}}+8\right]E_{p}\\
 & <\frac{\left|\nu\right|}{\left(h-z_{d,m}\right)^{2}}-\frac{\left|\nu\right|}{z_{d,m}^{2}}+\frac{4mg}{\pi\epsilon_{0}\kappa_{3}\left|\nu\right|}.
\end{align*}
 One notices that 
\[
\frac{\left|\gamma\left(b^{3}-a^{3}\right)-b^{3}\right|}{z_{d,m}^{3}}+\frac{\left|\gamma\left(b^{3}-a^{3}\right)-b^{3}\right|}{\left(h-z_{d,m}\right)^{3}}+8=\left|\psi\right|,
\]
 where $\psi$ is defined in Eq. (\ref{eq:psi}). Thus, I have 
\[
\left|\psi\right|E_{p}<\frac{\left|\nu\right|}{\left(h-z_{d,m}\right)^{2}}-\frac{\left|\nu\right|}{z_{d,m}^{2}}+\frac{4mg}{\pi\epsilon_{0}\kappa_{3}\left|\nu\right|}
\]
 or 
\[
E_{p}<\frac{1}{\left|\psi\right|}\left[\frac{\left|\nu\right|}{\left(h-z_{d,m}\right)^{2}}-\frac{\left|\nu\right|}{z_{d,m}^{2}}+\frac{4mg}{\pi\epsilon_{0}\kappa_{3}\left|\nu\right|}\right].
\]
 Notice that $\left|\nu\right|$ can be expressed as 
\[
\left|\nu\right|=\frac{\left|Q_{T}\right|}{4\pi\epsilon_{0}\kappa_{3}},
\]
 where Eq. (\ref{eq:QT-in-terms-of-nu}) has been used. Thus, $E_{p}$
becomes 
\begin{equation}
E_{p}<\frac{1}{\left|Q_{T}\psi\right|}\left\{ \frac{\left|Q_{T}\right|^{2}}{4\pi\epsilon_{0}\kappa_{3}}\left[\frac{1}{\left(h-z_{d,m}\right)^{2}}-\frac{1}{z_{d,m}^{2}}\right]+16mg\right\} .\label{eq:Ep-OSC-FINAL-nC}
\end{equation}
 Now, $E_{p}$ cannot be negative because it is the magnitude of electric
field. Since $z_{d,m}>h/2$ in Fig. \ref{fig:zd(t)-plot-nC}, it must
be true that 
\[
\frac{1}{\left(h-z_{d,m}\right)^{2}}>\frac{1}{z_{d,m}^{2}}.
\]
 Therefore, Eq. (\ref{eq:Ep-OSC-FINAL-nC}) can always be satisfied
for a core-shell structured particle which is sufficiently charged
negatively. Rearranging Eq. (\ref{eq:Ep-OSC-FINAL-nC}), I have 
\[
\frac{\left|Q_{T}\right|^{2}}{4\pi\epsilon_{0}\kappa_{3}}\left[\frac{1}{\left(h-z_{d,m}\right)^{2}}-\frac{1}{z_{d,m}^{2}}\right]-\left|Q_{T}\psi\right|E_{p}+16mg>0.
\]
 Introducing $\eta,$ 
\[
\eta=\frac{1}{\left(h-z_{d,m}\right)^{2}}-\frac{1}{z_{d,m}^{2}},
\]
 this becomes 
\[
\frac{\left|Q_{T}\right|^{2}\eta}{4\pi\epsilon_{0}\kappa_{3}}-\left|Q_{T}\psi\right|E_{p}+16mg>0.
\]
 Utilizing the quadratic formula, I find 
\[
\left|Q_{T}\right|>\frac{2\pi\epsilon_{0}\kappa_{3}}{\eta}\left(\left|\psi\right|E_{p}\pm\sqrt{\psi^{2}E_{p}^{2}-\frac{16mg\eta}{\pi\epsilon_{0}\kappa_{3}}}\right)
\]
 Since $\left|Q_{T}\right|$ cannot be negative, one must make sure
the right side is positive. But, since 
\[
\left|\psi\right|E_{p}>\sqrt{\psi^{2}E_{p}^{2}-\frac{16mg\eta}{\pi\epsilon_{0}\kappa_{3}}},
\]
 I have 
\begin{equation}
\left|Q_{T}\right|>\frac{2\pi\epsilon_{0}\kappa_{3}}{\eta}\left(\left|\psi\right|E_{p}-\sqrt{\psi^{2}E_{p}^{2}-\frac{16mg\eta}{\pi\epsilon_{0}\kappa_{3}}}\right),\label{eq:QT-OSC-FINAL-nC}
\end{equation}
 where 
\[
\eta=\frac{1}{\left(h-z_{d,m}\right)^{2}}-\frac{1}{z_{d,m}^{2}}.
\]
 Equation (\ref{eq:QT-OSC-FINAL-nC}) is the oscillation criterion
for the negatively charged core-shell structured particle. The resulting
form is different from the positively charged case, i.e., Eq. (\ref{eq:QT-OSC-FINAL}),
due to the fact that negatively charged particle oscillates near the
lower conductor plate whereas the positively charged particle oscillates
near the upper conductor plate in Fig. \ref{fig:particle-in-capacitor}.

\subsection{Electromagnetic radiation }

It is well known that the oscillating charged-particle radiates electromagnetic
energy. With respect to the reference point on the surface of the
upper conductor plate, the oscillating charged-particle has a dipole
moment given by 
\[
\mathbf{p}_{d}=-\mathbf{e}_{z}Q_{T}s
\]
 or 
\[
\mathbf{p}_{d}=-\mathbf{e}_{z}4\pi\epsilon_{0}\kappa_{3}\nu s
\]
 where $\mathbf{p}_{d}\equiv\mathbf{p}_{d}\left(t\right),$ $s\equiv s\left(t\right),$
and Eq. (\ref{eq:QT-in-terms-of-nu}) has been inserted for $Q_{T}.$
The negative sign comes from the fact that the particle is perceived
as residing in the negative $z\textnormal{-axis}$ to someone on the
surface of the upper conductor plate. In terms of $z_{d}\equiv z_{d}\left(t\right)$
defined in Eq. (\ref{eq:s-in-zd}), $\mathbf{p}_{d}$ becomes 
\begin{align*}
\mathbf{p}_{d} & =-\mathbf{e}_{z}4\pi\epsilon_{0}\kappa_{3}\nu\left(z_{d}+b\right),\\
\ddot{\mathbf{p}}_{d} & =-\mathbf{e}_{z}4\pi\epsilon_{0}\kappa_{3}\nu\ddot{z}_{d},
\end{align*}
 where $b$ is a constant. The electromagnetic power radiated by an
oscillating charged-particle, $P_{rad},$ is given by the Liénard
formula, 
\begin{align*}
P_{rad} & =\frac{1}{6\pi\epsilon_{0}c^{3}}\left(1-\frac{\dot{z}_{d}^{2}}{c^{2}}\right)^{-3}\ddot{\mathbf{p}}_{d}\cdot\ddot{\mathbf{p}}_{d}\\
 & =\frac{8\pi\epsilon_{0}\kappa_{3}^{2}\nu^{2}}{3c^{3}}\left(1-\frac{\dot{z}_{d}^{2}}{c^{2}}\right)^{-3}\ddot{z}_{d}^{2}.
\end{align*}
 Insertion of Eq. (\ref{eq:ODE-rela}) for $\ddot{z}_{d}$ finally
yields 
\begin{align*}
P_{rad} & =\frac{8\pi\epsilon_{0}\kappa_{3}^{2}\nu^{2}}{3c^{3}}\left(\frac{\pi\epsilon_{0}\kappa_{3}\nu}{4m}\left\{ \frac{\nu}{\left(z_{d}+b\right)^{2}}\right.\right.\\
 & -\frac{\nu}{\left(h-z_{d}-b\right)^{2}}+\frac{\left[\gamma\left(b^{3}-a^{3}\right)-b^{3}\right]E_{p}}{\left(z_{d}+b\right)^{3}}\\
 & \left.\left.+\frac{\left[\gamma\left(b^{3}-a^{3}\right)-b^{3}\right]E_{p}}{\left(h-z_{d}-b\right)^{3}}-8E_{p}\right\} -g\right)^{2},
\end{align*}
 where 
\begin{align*}
\nu & =\frac{2a\left(b-a\right)\sigma_{1}}{\epsilon_{0}\kappa_{2}}+\frac{a^{2}\sigma_{1}+b^{2}\sigma_{2}}{\epsilon_{0}\kappa_{3}},
\end{align*}
 which is the result defined in Eq. (\ref{eq:radiation-power}) for
the Liénard radiation power.

\section{Concluding Remarks }

The phenomenon of charged-particle oscillation subjected to a constant
electric field has been investigated. For a positively charged core-shell
structured particle, the criterion for an oscillatory motion is given
by 
\begin{equation}
Q_{T}>\frac{2\pi\epsilon_{0}\kappa_{3}}{\xi}\left(\left|\psi\right|E_{p}+\sqrt{\psi^{2}E_{p}^{2}+\frac{16mg\xi}{\pi\epsilon_{0}\kappa_{3}}}\right),\label{eq:OCR}
\end{equation}
 where 
\[
\xi=\frac{1}{z_{d,m}^{2}}-\frac{1}{\left(h-z_{d,m}\right)^{2}}>0,
\]
 
\[
\psi=\frac{\gamma\left(b^{3}-a^{3}\right)-b^{3}}{z_{d,m}^{3}}+\frac{\gamma\left(b^{3}-a^{3}\right)-b^{3}}{\left(h-z_{d,m}\right)^{3}}-8<0,
\]
 and 
\[
\gamma=\frac{3\kappa_{3}b^{3}}{\left(\kappa_{2}+2\kappa_{3}\right)b^{3}+2\left(\kappa_{2}-\kappa_{3}\right)a^{3}}<1.
\]
 For the parallel plate configuration illustrated in Fig. \ref{fig:particle-in-capacitor},
the positively charged core-shell structured particle can only have
oscillatory modes in region where $0<z_{d}<h/2.$ This region is further
divided into subregions $A$ and $B,$ and this is illustrated in
Fig. \ref{fig:mechanism}. In this work, the upper conductor plate
is located at $z_{d}=0$ and the lower conductor plate is located
at $z_{d}=h.$ Further, the upper conductor plate is held at voltage
$V_{T}$ and the lower conductor plate has a voltage of $V_{L},$
where $V_{T}>V_{L}.$ That said, the magnitude of a dominant force
in region $A$ falls off with distance like $1/s^{3}$ and this force
acts to repulse the positively charged core-shell structured particle
from the upper conductor plate. In region $B,$ the magnitude of a
dominant force falls of like $1/s^{2}$ with distance and this force
acts to attract the particle towards the upper conductor plate. It
is this competition between the two dominant forces from regions $A$
and $B$ that puts particle in an oscillatory motion. 

Such oscillatory mechanism does not involve any charge exchange; therefore,
it is fundamentally different from the traditional description by
charge exchange.\cite{Tulagin} Nevertheless, the novel finding in
this work does not invalidate the charge exchange description altogether
because the two oscillatory phenomena are not exactly the same. For
instance, the traditional picture deals with charged particle oscillation
in which the particle sweeps the entire gap between the two plates
whereas, here, the particle only oscillates in the restricted regions
between the plates. Further, here, the charged-particle must be structured
and not a ``point'' particle in order to have any oscillatory motion.
The reason for this is because the repulsive force in region $A$
arises as a consequence of induced polarization and a point particle
does not have such property. Without such force appearing in region
$A,$ there would not be a way to repulse the positively charged particle
from sticking to the surface of the upper conductor plate in Fig.
\ref{fig:mechanism}. The traditional description by process of charge
transfer does not have such restrictions, of course. Thus, the two
phenomena are not exactly the same. Nonetheless, the novel finding
here presents yet another mechanism for a charged-particle oscillation
in an otherwise uniform and constant electric field. 

For a negatively charged core-shell structured particle, the oscillatory
criterion is given by 
\begin{equation}
\left|Q_{T}\right|>\frac{2\pi\epsilon_{0}\kappa_{3}}{\eta}\left(\left|\psi\right|E_{p}-\sqrt{\psi^{2}E_{p}^{2}-\frac{16mg\eta}{\pi\epsilon_{0}\kappa_{3}}}\right),\label{eq:OCR-nC}
\end{equation}
 where 
\[
\eta=\frac{1}{\left(h-z_{d,m}\right)^{2}}-\frac{1}{z_{d,m}^{2}}>0.
\]
 The oscillatory behavior for negatively charged core-shell structured
particle is observed only in the region where $h/2<z_{d}<h;$ and,
this is illustrated in Fig. \ref{fig:mechanism-nC}. For the case
of negatively charged particle, subregions $A$ and $B$ are formed
near the lower conductor plate side of Fig. \ref{fig:mechanism-nC},
which is exactly the opposite of the positively charged particle case.
Again, the force in region $A$ repulses the particle from the lower
conductor plate and the force in region $B$ attracts the particle
to the lower conductor plate. It is this ``push-pull'' competition
between the two dominant forces in regions $A$ and $B$ that puts
charged-particle in an oscillatory motion. Such oscillatory behavior
is only possible because the particle can be polarized under applied
electric field. 

Because of the explicit mass dependence in Eqs. (\ref{eq:OCR}) and
(\ref{eq:OCR-nC}), the charged particle oscillation discussed in
this work is more favorably satisfied by microscopic or smaller particles
than by macroscopic counterparts due to various experimental limitations.
In principle, particles of any size can be charged to satisfy the
oscillation criterion of Eqs. (\ref{eq:OCR}) or (\ref{eq:OCR-nC})
provided a DC electric field of sufficient strength can be applied
without electrical breakdown taking place. Unfortunately, electrical
breakdown occurs at some point even in vacuum and this may limit macroscopic
particles from satisfying Eqs. (\ref{eq:OCR}) or (\ref{eq:OCR-nC}).\cite{IEEE-vacuum} 

The charged-particle oscillator based on the presented novel mechanism
represents a natural prototype for illuminating electric dipole radiation.
In such system, the frequency of emitted electromagnetic radiation
is controlled by a DC voltage biased across the two plane-parallel
electrodes. The strength of emitted radiation power from such system
depends directly on the magnitude of effective charge carried by the
charged-particle. 

As for potential applications, the finding in this work can be utilized
to build a source for generating microwave radiation. Microwaves thus
generated, for instance, might be used to heat water or to excite
gases in tiny plasma capsules to produce light. As another potential
application, the repulsive mechanism discussed here can be utilized
to build an anti-friction device. Such device would not require any
grease or oil, which are known to be environmentally hazardous. Instead,
the friction in such device would be controlled by applied DC voltage.

In nanoscale plasma confinements, nanocavities for instance, only
scarce number of ionized atoms or particles participate in the dynamics.
This is so because only finite number of ionized particles can be
contained in such small scale confinement. In such system, effects
arising from the collective motion of particles may only be a small
perturbation compared to the effects arising from individual particle
dynamics. For very large scale systems, wherein enormous number of
ionized particles are involved, exactly the opposite is true. There,
effects arising from the collective motion of particles become predominant.
As the technology evolves, plasma systems will eventually enter the
regime of submicron to nanometer scale plasma confinements. Already,
the size of display pixels is approaching the dimensions of few visible
wavelengths; and, display technologies based on charged particles
are beginning to involve fewer ionized nanoparticles, where individual
particle effects are no longer small perturbations. In the electronics
industry, the finite particle nonneutral plasma systems are becoming
technologically very important as nanoscale fabrication processes
demand for the development of extreme UV (ultraviolet) to X-ray lasers.
For instance, device fabrication process at length scale of $\sim20\,\textnormal{nm}$
or less by photolithography requires extreme UV lasers. Because solid
state devices cannot generate such high frequency laser waves, plasma
based sources are the only viable candidates for building extreme
UV or X-ray lasers. The macroscopic plasma sources can readily generate
extreme UV electromagnetic waves, but at the cost of losing the coherent
nature of laser waves. Inevitably, the coherent nature of a laser
source demands for finite particle plasma sources, wherein the physics
of individual particle dynamics is predominantly important. In this
respect, the results obtained in this work should be useful and interesting
to certain areas of plasma physics.

\section{Acknowledgments}

The author acknowledges the support for this work provided by Samsung
Electronics Co., Ltd.

\end{document}